\documentclass[%
    aps,
    prd,
    10pt,
    letterpaper,
    superscriptaddress,
    preprintnumbers,
    floatfix,
    twocolumn,
    nofootinbib,
    showpacs,
    hyphens
]{revtex4-2}  
\usepackage{amsmath}
\usepackage{amsfonts}
\usepackage{mathtools}
\usepackage{cases}
\usepackage{lmodern}
\usepackage[T1]{fontenc}
\usepackage{upgreek}
\usepackage{microtype}
\usepackage{graphicx} 
\usepackage{comment}    
\usepackage{dcolumn}   
\usepackage{bm}      
\usepackage[shortcuts, cyremdash]{extdash}
\newcommand{\myshortndash}{\scalebox{0.82}{--}}
\usepackage[caption=false,justification=justified,font=normalsize,labelfont=normalsize]{subfig}
\usepackage{ragged2e} %
\DeclareCaptionJustification{justified}{\justifying}
\usepackage{caption}
\usepackage{multirow}
\usepackage{slashed}
\usepackage{placeins}
\usepackage{epstopdf}
\usepackage{color}
\usepackage{tikz,pgfplots}
\pgfplotsset{compat=newest}
\usepackage{enumerate,comment}
\usepackage{enumitem}
\usepackage{setspace}
\usepackage{dsfont}
\usepackage{nicefrac}
\usepackage{siunitx}
\usepackage{physics}
\usepackage[normalem]{ulem}
\usepackage{tensor}
\usepackage{tabularx}
\usepackage{listings}
\usepackage{adjustbox}
\usepackage{hyperref}  
\hypersetup{breaklinks=true}
\usepackage[capitalize]{cleveref}

\crefname{section}{Sec.}{Secs.}
\Crefname{section}{Sec.}{Secs.}
\definecolor{c1}{RGB}{206,0,0}
\definecolor{c2}{RGB}{249,149,0}
\definecolor{c3}{RGB}{153,0,210}
\definecolor{c4}{RGB}{0,109,219}
\definecolor{c5}{RGB}{0,146,146}
\definecolor{c6}{RGB}{255,109,182}
\pgfplotscreateplotcyclelist{cycle1}{
{c1}, {c2}, {c3}, {c4}, {c5}, {c6}
}

\begin{document}
\title{First constraints on the nonperturbative gluon Collins-Soper kernel}
\author{Artur Avkhadiev}
\affiliation{Physics Division, Argonne National Laboratory, Lemont, IL 60439, USA}
\affiliation{Center for Theoretical Physics\Emdash{}a Leinweber Institute, Massachusetts Institute of Technology, Cambridge, MA 02139, U.S.A.}
\author{Yang Fu}
\affiliation{Center for Theoretical Physics\Emdash{}a Leinweber Institute, Massachusetts Institute of Technology, Cambridge, MA 02139, U.S.A.}
\author{Phiala E. Shanahan}
\affiliation{Center for Theoretical Physics\Emdash{}a Leinweber Institute, Massachusetts Institute of Technology, Cambridge, MA 02139, U.S.A.}
\author{Michael L. Wagman}
 \affiliation{Fermi National Accelerator Laboratory, Batavia, IL 60510, USA}  
\author{Yong Zhao}
\affiliation{Physics Division, Argonne National Laboratory, Lemont, IL 60439, USA}
\preprint{FERMILAB-PUB-26-0391-T, MIT-CTP/6053, INT-PUB-26-024}
\newcommand{\Min}{{\small \mathrm{M}}} %
\newcommand{\Euc}{{\small \mathrm{Euc.}}} %
\newcommand{\mom}[1]{\mathbf{#1}}
\newcommand{\pos}[1]{\mathbf{#1}}
\newcommand{\tran}{T} %
\newcommand{\wf}{\tilde{\phi}}
\newcommand{\symm}{{\small \textrm{symm.}}}
\newcommand{\qcntr}{{\small q\textrm{-cent.}}}
\newcommand{\unsubtr}{{\small \mathrm{unsub.}}} %
\newcommand{\bare}{{\small \mathrm{bare}}} 
\newcommand{\unexp}{{\mathrm{u}}} 
\newcommand{\ren}{{\small \mathrm{ren.}}} %
\newcommand{\nn}{\nonumber} %
\DeclareRobustCommand{\Eq}[1]{Eq.~\eqref{eq:#1}}
\DeclareRobustCommand{\Eqs}[2]{Eqs.~\eqref{eq:#1} and \eqref{eq:#2}}
\DeclareRobustCommand{\fig}[1]{Fig.~\ref{fig:#1}}
\DeclareRobustCommand{\figs}[2]{Figs.~\ref{fig:#1} and \ref{fig:#2}}
\DeclareRobustCommand{\app}[1]{App.~\ref{app:#1}}
\DeclareRobustCommand{\sec}[1]{Sec.~\ref{sec:#1}}
\DeclareRobustCommand{\secs}[2]{Secs.~\ref{sec:#1} and \ref{sec:#2}}
\DeclareRobustCommand{\tbl}[1]{Table~\ref{tbl:#1}}
\DeclareRobustCommand{\refcite}[1]{Ref.~\cite{#1}}
\DeclareRobustCommand{\refcites}[1]{Refs.~\cite{#1}}
\newcommand{\MSbar}{
\overline{\rm MS}    
} 
 \newcommand{\xMOM}{ 
 \rm RI/xMOM
 } 
\newcommand{\MOM}{
    \rm RI^\prime/MOM
 }

\newcommand{\DNP}{\mathcal{D}_{\text{NP}}}
\newcommand{\DP}{\mathcal{D}_\text{res}}

\begin{abstract}
The gluon Collins-Soper kernel, which encodes the rapidity evolution of transverse-momentum-dependent gluon distributions, is constrained for the first time in the nonperturbative regime, for transverse momentum scales $q_{T} \in [ \SI{300}{\MeV}, \SI{1.3}{\GeV}]$.
The constraints are determined in lattice QCD at a close-to-physical pion mass $M_\pi = \SI{172(3)}{\MeV}$, a single lattice spacing $a=\SI{0.15}{\femto\meter}$, and next-to-next-to-leading logarithmic matching in Large-Momentum Effective Theory. 
These results represent the first step toward a controlled determination of the gluon Collins-Soper kernel in QCD, with eventual phenomenological import and relevance to present and future experiments sensitive to the gluon structure of hadronic matter.
\end{abstract}
\maketitle
\par Resolving the multidimensional gluon structure of hadronic matter is a key objective of numerous ongoing, planned, and proposed experimental programs in particle and nuclear physics~\cite{Aschenauer:2015eha,RHICSPIN:2023zxx,%RHIC,BNL
Dudek:2012vr,CLAS:2021opg,%CEBAF,JeffLab
Accardi:2012qut,AbdulKhalek:2022hcn,%EIC,BNL
LHeC:2020van,%LHeC,CERN
Anderle:2021wcy%EiCC
}.
One aspect of this structure is the intrinsic transverse motion of gluons in hadrons boosted close to the speed of light.
Such motion, characterized by transverse momentum $q_T$, can be described in QCD in terms of gluon transverse-momentum-dependent (TMD) partonic functions, or gluon TMDs~\cite{Mulders:2000sh} (see Refs.~\cite{Diehl:2015uka,Boussarie:2023izj} for reviews).
Descriptions of experimental observables based on gluon TMDs are universal across semi-inclusive deep inelastic scattering~\cite{Boer:2010zf,Kang:2020xgk,Echevarria:2026vca,Kishore:2022ddb} and several processes in hadron\myshortndash{}hadron collisions including photoproduction in ultraperipheral collisions~\cite{Nadolsky:2007ba,Qiu:2011ai,Anedda:2025bts}, quarkonium production in color-singlet channels~\cite{Boer:2012bt,Echevarria:2019ynx,denDunnen:2014kjo,Lansberg:2017tlc,Kato:2024vzt,Scarpa:2019fol,Chakrabarti:2022rjr}, and Higgs production~\cite{Catani:2010pd,Echevarria:2015uaa,Anedda:2026cox}.
\par Constraining TMDs experimentally requires a simultaneous extraction of quark and gluon Collins-Soper (CS) kernels~\cite{Collins:1981va,Collins:1981uk,Collins:1984kg}: additional nonperturbative functions relating TMDs across rapidity scales.
While the quark kernel has been nonperturbatively constrained both phenomenologically\Emdash{}in global fits~\cite{Landry:2002ix,%BLNY
Moos:2025sal,%ART25
Bacchetta:2025ara,%MAPNN
Avkhadiev:2025wps,%MAP+Lat25
Kang:2024dja,%EEC24
Camarda:2025lbt%CFR25
}
and via a novel approach based on parton-shower kinematics~\cite{Martinez:2024mou}\Emdash{}and from first principles using lattice QCD~\cite{
% --- MIT / FNAL / ANL
Shanahan:2020zxr,% SWZ20
Shanahan:2021tst,%SWZ21
Avkhadiev:2023poz,%ASWZ23
Avkhadiev:2024mgd,%ASWZ24
% LPC ---
LatticeParton:2020uhz,%LPC20
LPC:2022ibr,%LPC22
Chu:2023flm,%LPC23
Tan:2025ofx,%LPC25
% --- Regensburg  ---
Schlemmer:2021aij,%SVZES21
Shu:2023cot,
% --- ETMC / PKU ---
Li:2021wvl,% PKU/ETMC21
Alexandrou:2025xci,
% --- BNL ---
Bollweg:2024zet,%DWF24
Bollweg:2025iol,%PionCG25
% --- Wilson lines ---
Francis:2026czb}, no nonperturbative constraints to date have been made on the gluon kernel.
At present, the only available constraints on this quantity are from perturbative QCD, where, to fourth loop order~\cite{vonManteuffel:2020vjv,Moult:2022xzt,Duhr:2022yyp}, the gluon kernel is related to the quark kernel by Casimir scaling: the ratio $C_{\text{A}}/C_{\text{F}}$ of the adjoint- to fundamental-representation quadratic Casimir $SU(3)$ invariants~\cite{Bali:2000un}.
It is not known whether or not this relation holds in the nonperturbative regime.
In the absence of nonperturbative constraints, however, Casimir scaling of the CS kernels guides phenomenology: for example, it was recently applied in the first phenomenological extraction of unpolarized gluon TMDs~\cite{Anedda:2026cox}.
\par Systematically controlled lattice QCD constraints on the gluon CS kernel could guide both phenomenology and theory~\cite{Moult:2025nhu,Cridge:2025wwo}.
For analogous constraints on the quark kernel~\cite{Avkhadiev:2023poz,Avkhadiev:2024mgd}, this has already been demonstrated: in joint global TMD fits to experimental data and numerical lattice QCD results~\cite{Avkhadiev:2025wps}, in extracting energy\myshortndash{}energy correlators in the back-to-back limit of $e^+ e^-$ collisions~\cite{Jaarsma:2025tck},
and in quantifying nonperturbative QCD uncertainties in the measurement of the $W$-boson mass~\cite{CMS:2024lrd}.
\par This Letter presents the first constraints on the nonperturbative gluon CS kernel, obtained in the range of transverse momentum scales $q_T  \in [ \SI{300}{\MeV}, \SI{1.3}{\GeV}]$ from lattice QCD at a single lattice spacing.
The obtained constraints are compatible within $1$\myshortndash{}$2\sigma$ with the Casimir scaling of the quark CS kernel, pave the way for future determinations within the same approach, and may inform future phenomenological analyses of three-dimensional hadron structure.
\par {\textbf{Theoretical framework:}} The gluon CS kernel can be derived from any gluon TMD as a universal anomalous dimension.
In the calculation below, it is extracted as
\begin{align}
\label{eq:cs-kernel-derivative}
       \gamma_{g}(b_T, \mu)
        = \dv{\ln \sqrt{\zeta}} \ln B_{g/h}\left(x,b_T,\mu,\zeta{/\nu^2}\right)\thinspace,
\end{align}
where the chosen TMD is a gluon TMD beam function $B_{g/h}  (x,b_T,\mu,\zeta{/\nu^2})$~\cite{Stewart:2009yx,Jain:2011iu,Boussarie:2023izj}: the collinear component of a gluon TMD parton distribution function encoding a number density of gluons $g$ in an ultrarelativistic hadron $h$.
Here, $x$ and $b_T$ respectively denote collinear momentum fraction and transverse displacement Fourier-conjugate to $q_T$. %
The scales $\mu$ and $\zeta$ respectively govern virtuality and rapidity (CS) evolution~\cite{Collins:1981va,Collins:1981uk,Collins:1984kg}, while the auxiliary rapidity renormalization scale $\nu$ does not affect the definition of the kernel~\cite{Chiu:2011qc,Chiu:2012ir}. %
The nonperturbative regime in $\gamma_{g}(b_T, \mu)$ is $b_T \gtrsim \Lambda_{\rm QCD}^{-1}$, independent of $\mu$.
\par A direct evaluation of Eq.~\eqref{eq:cs-kernel-derivative} with lattice QCD  is not possible, because TMDs are defined through lightlike-separated operators whose matrix elements are inaccessible in Euclidean space.
Instead, it is possible to determine bare quasi-TMD beam functions~\cite{Ji:2013dva,Ji:2014hxa,Ebert:2018gzl}, defined as Fourier transforms 
\begin{equation}
\label{eq:ft}
\begin{aligned}
         & \tilde{B}_{g/h}\left(x,b_T,a, P_z\right) 
        \\ &\quad= \frac{1}{x}
        \int
        \frac{{{P_z \dd b^z}}}{2\pi} e^{i x P_z b^z}
        \lim_{\ell/a \to \infty} 
        \tilde{B}^{\ell}_{g/h}(b^z, b_T, a, P_z)\thinspace,
\end{aligned}
\end{equation}
where $a$ denotes the lattice spacing, $P_z$ denotes the momentum of the hadron boosted to $\mom{P} = P_z \mathbf{\hat{z}}$, and $\ell$ parametrizes the finite extent of spacelike-separated operators in matrix elements that are accessible to Euclidean lattice QCD. These matrix elements are computed as
\begin{equation}
\label{eq:matrix-element}
\begin{aligned}
\tilde{B}^{\ell}_{g/h}(b^z, b_T, a, P_z) &=
    N_{g/h}(a, P_z)
    \frac{\tilde{\Omega}^{\ell}_{g/h}(b^z, b_T, a, P_z)}{[{{\tilde{\mathcal{Z}}^{2\ell}_{g}(b_T, a)}}]^{1/2}}\thinspace
\end{aligned}
\end{equation}
where 
\begin{equation}
\label{eq:tmd-correlator}
    \tilde{\Omega}^{\ell}_{g/h}(b^z, b_T, a, P_z) = \ev{O^{\ell}_g(y, b^z, \pos{b}_T)}{h(P_z)}\thinspace,
\end{equation}
is a matrix element in a relativistically normalized hadron state  $\ket{h(P_z)}$  and $O^{\ell}_g(y, b^z, \pos{b}_T)$ denotes a nonlocal operator with spacelike separation $b^z \pos{\hat{z}} + \pos{b}_T$ ($b_T = |\pos{b}_T|$);
it is defined further below alongside the normalization factor $N_{g/h}(a, P_z)$ and the auxiliary matrix element $\tilde{\mathcal{Z}}^{2\ell}_{g}(b_T, a)$.
\par An evaluation of Eq.~\eqref{eq:cs-kernel-derivative} via quasi-TMD beam functions~\cite{Ji:2014hxa,Ebert:2018gzl} is then enabled by Large-Momentum Effective Theory (LaMET)~\cite{Ji:2013dva,Ji:2014gla,Ji:2020ect} for $x P_z~\gg~\{ \Lambda, b^{-1}_T\}$ where $\Lambda$ denotes a generic hadron scale:
\begin{equation}
\label{eq:cs-kernel-derivative-matched}
\begin{aligned}
    \gamma_{g}(b_T, \mu)
        &= \lim_{a\to 0} \dv{\ln P_z} \Big\lbrack
         \ln {\tilde{B}^{}_{g/h}\left(x,b_T,a,P_z\right)}
         \\ &\qquad\qquad\qquad\quad- 
         \ln H_g(\mu, x P_z) \Big\rbrack+ \text{p.c.}
\end{aligned}
\end{equation}
Here, $H_{g}(\mu, x P_z)$ denotes a matching kernel calculated perturbatively in LaMET~\cite{Schindler:2022eva,Zhu:2022bja}, with power corrections
($\text{``p.c.''}$) in $\Lambda/(x P_z)$ and $1/(b_T x P_z)$. 
In the limit $P_z \to \infty$, the corrections vanish and several possible definitions of $\tilde{B}_{g/h}\left(x,b_T,a,P_z\right)$ are equivalent. 
This calculation adopts a definition permitting multiplicative, $P_z$-independent renormalization with a subsequent conversion to the $\MSbar$ scheme~\cite{Zhang:2018diq,Li:2018tpe}. 
Given this choice, renormalization constants do not appear in Eq.~\eqref{eq:cs-kernel-derivative-matched}, and dependence on $\mu$ is carried solely by $H_{g}(\mu, x P_z)$.
\par {\it {Nonperturbative matrix elements.}}\Emdash{}The operator $O^{\ell}_g(y, b^z, \pos{b}_T)$ in Eq.~\eqref{eq:tmd-correlator} is defined by a trace-subtracted combination
\begin{equation}
\label{eq:operator-0i0i}
\begin{aligned}
&O_g^{\ell}(y, b^z, \pos{b}_T) 
=  \tfrac{1}{2}\sum_{\mathclap{\alpha,\beta \in \lbrace x,y \rbrace}}\Big\lbrack O_g^{tt\alpha\beta,\ell}(y, b^z, \pos{b}_T) 
\\ &\qquad\qquad\qquad\qquad\qquad- \tfrac{1}{3}\sum_{\mathclap{\rho,\sigma \in \lbrace x,y,t \rbrace}} O_g^{\rho\sigma\alpha\beta,\ell}(y, b^z, \pos{b}_T)\Big\rbrack\thinspace
\end{aligned}
\end{equation}
of the operator $O_g^{\mu\nu\alpha\beta,\ell}(y, b^z, \pos{b}_T)$ composed of a pair of gluon field strength tensors connected by Wilson lines~\cite{Boussarie:2023izj}:
\begin{equation}
\label{eq:operator-0i0i-detail}
\begin{aligned}
&O_g^{\mu\nu\alpha\beta,\ell}(y, b^z, \pos{b}_T) = 2 \Tr\big[F^{\mu\alpha}(y+b) W^{\ell}_{\sqsupset}(y, b^z, \pos{b}_T) 
\\&\qquad\qquad\qquad\qquad\qquad\quad\times F^{\nu\beta}(y) W^{-\ell}_{\sqsupset}(y, b^z, \pos{b}_T) \big]\thinspace.
\end{aligned}
\end{equation}
Here, $\mu,\nu$ and $\alpha,\beta$ respectively correspond to longitudinal and transverse directions,  $F^{\mu\alpha}(y)$ denotes a component of the gluon field strength tensor, and $W^{\pm \ell}_{\sqsupset}(y, b^z, \pos{b}_T)$ denotes a fundamental Wilson line of length $\ell + b_T$ comprising a staple-shaped combination of segments~\cite{Zhang:2018diq}.
The corresponding normalization factor in Eq.~\eqref{eq:matrix-element} for the hadron state $\ket{h(P_z)}$ is given by
\begin{equation}
\label{eq:normalization}
    N_{g/h}(a, P_z)  = \frac{1}{\tfrac{2}{3} (E^h_{P_z})^{2}}\thinspace,
\end{equation} 
where $E^h_{P_z}$ denotes the energy of the boosted hadron state, with $a$-dependence suppressed.
\par The definition of $\tilde{\mathcal{Z}}^{2\ell}_g(b_T, a)$ in Eq.~\eqref{eq:matrix-element} is scheme-dependent and does not affect $\gamma_{g}(b_T, \mu)$, only serving to make the limit $\ell/a \to \infty$ in Eq.~\eqref{eq:ft} well-defined by subtracting the leading divergences in $O_g^{\ell}(y, b^z, \pos{b}_T)$, which are logarithmic in $a$ and linear in $\ell + b_T$~\cite{Zhang:2018diq}.
In this work, $\tilde{\mathcal{Z}}^{2\ell}_g(b_T, a)$ is given by a rectangular loop of Wilson lines~\cite{Ji:2018hvs,Zhang:2022xuw} in the adjoint representation with total length $2(\ell + b_T)$, and can be expressed as
\begin{equation}
\begin{aligned}
\label{eq:matrix-element-loop}
    \tilde{\mathcal{Z}}^{2\ell}_g(b_T, a) &=
    \tfrac{1}{N_c^2 - 1}\ev{\left|\mathrm{Tr}\left[W^{2\ell}_{\square}(\pos{b}_T)\right]\right|^2 - 1}{0}\thinspace,
\end{aligned}
\end{equation}
where $N_\text{c}=3$ and $W^{2\ell}_{\square}(\pos{b}_T)$ denotes the corresponding Wilson loop in the fundamental representation~\cite{Dorn:1981wa}. 
\par {\it {Perturbative matching.}}\Emdash{}The matching in Eq.~\eqref{eq:cs-kernel-derivative-matched} is calculated at next-to-next-to-leading logarithmic (NNLL) accuracy and extended to a $b_{T}$-dependent, or ``$b_{T}$-unexpanded'' (uNNLL) form, with the kernels denoted $H^{\text{NNLL}}_g(x, \mu, P_z)$ and $H^{\text{uNNLL}}_g(x, b_T, \mu, P_z)$, respectively.
To calculate the NNLL kernel, fixed-order kernels are adapted from Ref.~\cite{Zhu:2022bja} for the operator in Eq.~\eqref{eq:operator-0i0i}, with logarithms resummed from an initial scale of $\mu_0 = 2 x P_z$ to $\mu = \SI{2}{\GeV}$ in the $\overline{\text{MS}}$ scheme following Refs.~\cite{Ji:2019ewn,Ebert:2022fmh} and using resummation kernels in Refs.~\cite{Gaunt:2014cfa,Luo:2019bmw}.
The uNNLL kernel additionally incorporates the $b_T$-dependent corrections in Eq.~\eqref{eq:cs-kernel-derivative-matched} while preserving multiplicative matching form as detailed in Ref.~\cite{Avkhadiev:2023poz}.
Kernel definitions are provided in the Supplementary Material.
\par \textbf{Lattice QCD calculation:} Constraints on $\gamma_{g}(b_T, \mu)$ based on Eqs.~\eqref{eq:ft}\myshortndash{}\eqref{eq:matrix-element-loop} are extracted with lattice QCD, with each stage of the calculation summarized below.
Additional figures and details are included in the Supplementary Material.
\par {\it {Computational setup}.}\Emdash{}The analysis is performed using an ensemble of $1105$ gauge field configurations characterized by space-time volume $L^3 \times T = a^4(32^3 \times 48)$ with $a=\SI{0.15}{\femto\meter}$.
This ensemble is produced by the MILC collaboration~\cite{MILC:2012znn} using a one-loop Symanzik-improved gauge action~\cite{Symanzik:1983dc,Luscher:1984xn}, a highly-improved staggered quark action with $2+1+1$ dynamical flavors, and with sea quark masses tuned to reproduce the physical masses of the pion and the kaon~\cite{Follana:2006rc}.
To enhance the signal-to-noise ratio in numerical results, the gauge-field configurations are subjected to gradient (Wilson) flow~\cite{Luscher:2010iy}: with flow-time $\mathfrak{t}/a^2 = 2.0$ for calculations of gluon operators and Wilson loops, and with flow-time $\mathfrak{t}/a^2 = 1.0$ for calculations of two-point correlation functions.
The two-point functions are computed in a mixed-action setup using the tree-level Wilson-clover fermion action~\cite{Sheikholeslami:1985ij} with hopping parameter $\kappa=0.12575$ and a clover-term coefficient $c_{\text{sw}}=1.0$, resulting in a close-to-physical pion mass of $M_\pi = \SI{172\pm 3}{\MeV}$.
\par {\it {Extraction of matrix elements}.}\Emdash{}Using the gauge field ensemble, requisite matrix elements and energies are extracted for boosted pion states, $h(P_z) = \pi(P_z)$,
$\tilde{\Omega}^{\ell}_{g/\pi}(b^z, b_T, a, P_z)$ and $\tilde{\mathcal{Z}}^{2\ell}_g(b_T, a)$ are extracted with Wilson-line geometries characterized by $b_T / a \in \lbrace 1,2,3,4 \rbrace $, $|b^z| \leq \ell$ at each fixed $\ell$, and $\ell / a \in \lbrace 4,5,\ldots, \tfrac{1}{2a} L \rbrace$. 
Extractions of
$\tilde{\Omega}^{\ell}_{g/\pi}(b^z, b_T, a, P_z)$ and $E^\pi_{P_z}$ are performed for a range of pion momenta $P_z = 2\pi n_z / L$ given by  $4 \leq n_z \leq 8$, corresponding to $\SI{1.03}{\GeV} \lesssim P_z \lesssim \SI{2.05}{\GeV}$.
Pion states are created using interpolating operators 
\begin{equation}
    \chi_{\pi}^\dagger(\mathbf{x}, t) = \bar{u}_{{\mom{K}}}(\mathbf{x}, t) \gamma_4\gamma_5 d_{-{\mom{K}}}(\mathbf{x}, t)\thinspace,
\end{equation}
which comprise a kinematically enhanced Dirac-matrix structure~\cite{Zhang:2025hyo} and quark fields subjected to $32$ iterations of Gaussian momentum smearing~\cite{Bali:2016lva} with smearing width $\varepsilon = 0.2$, and with smearing momentum $\mom{K} = K_z \hat{\mathbf{z}}$ fixed to $K_z = \SI{1.16}{\GeV}$ for all $P_z$ corresponding to variable quark smearing momentum fractions in the range $K_z / P_z \in [0.56, 1.50]$. 
(The dependence on the smearing parameters in $\chi_\pi^\dagger(\pos{x}, t)$ is notationally suppressed.)
\par To obtain $\tilde{\Omega}^{\ell}_{g/\pi}(b^z, b_T, a, P_z)$ and $E^\pi_{P_z}$, two-point (``\text{2pt}'') and three-point (``\text{3pt}'') correlation functions in Euclidean time are computed according to
\begin{equation}
\label{eq:2pt}
\begin{aligned}
    C_{\text{2pt}}(t, a, \mom{P})
    &= a^6 \sum_{\mathclap{\pos{x}}} e^{ i \mom{P} \cdot \pos{x}} \langle \chi_{\pi}(\mathbf{x}, t) \chi_{\pi}^\dagger(\mathbf{0},0)\rangle\thinspace
\end{aligned}
\end{equation}
and 
\begin{align}
    &C^{\ell}_{\text{3pt}}(t, \tau, b^z, b_T, a, \mom{P} = P_z \mathbf{\hat{z}})
    \nonumber \\ \label{eq:3pt} &\begin{aligned}&\quad= \langle C_{\text{2pt}}(t, a, \mom{P}) \sum_{\mathclap{\pos{y}}} a^4 O_g^\ell((\mathbf{y}, \tau), b^z, b_T) 
    \\&\qquad\qquad- \langle C_{\text{2pt}}(t, a, \mom{P}) 
\rangle \langle \sum_{\mathclap{\pos{y}}} a^4 O_g^\ell((\mathbf{y}, \tau), b^z, b_T)
    \rangle 
    \end{aligned}
    \\ \label{eq:3pt-asymptote} &\quad\xrightarrow{t \gg \tau \gg a}  \tilde{\Omega}^{\ell}_{g/\pi}(b^z, b_T, a, P_z) \frac{|Z_\pi(P_z)|}{\left(2 E^\pi_{P_z}\right)^2} e^{-E^\pi_{P_z} t} + \ldots\thinspace
\end{align}
Here, $Z_\pi(P_z)$ denotes the overlap factor for the pion state (with dependence on the interpolating operator notationally suppressed), and the ellipsis denotes exponentially-suppressed corrections.
\par Given the correlation functions, $E^\pi_{P_z}$ and $\tilde{\Omega}^{\ell}_{g/\pi}(b^z, b_T, a, P_z)/(2E^\pi_{P_z})$ are extracted using the Lanczos algorithm~\cite{Wagman:2024rid,Hackett:2024xnx,Ostmeyer:2024qgu,Chakraborty:2024exj,Hackett:2024nbe,Abbott:2025yhm} with central values and statistical uncertainties estimated using correlated nested bootstrap samples; see Supplementary Material and Refs.~\cite{Hackett:2024xnx,Abbott:2025yhm} for details of matrix element determinations.
Results obtained from this procedure are confirmed to be consistent within uncertainties with statistical fits of $C_{\text{2pt}}(t, a, \mom{P})$ and ratios of three-point to two-point functions across a range of fitting windows in $t$ and $\tau$. 
To obtain $\tilde{\mathcal{Z}}^{2\ell}_g(b_T, a)$, Wilson loops $W^{2\ell}_{\square}(\pos{b}_T)$ are computed and combined according to Eq.~\eqref{eq:matrix-element-loop}, with  uncertainties determined identically to those of $\tilde{\Omega}^{\ell}_{g/\pi}(b^z, b_T, a, P_z)$ and $E^\pi_{P_z}$.
\begin{figure}[t]
\centering
\subfloat[{$\tilde{B}^{\ell}_{g/\pi}(b^z, b_T, a, P_z)$ as functions of $1/\ell$ at fixed $b^z$ (grayscale markers), and the corresponding $\tilde{B}_{g/\pi}(b^z, b_T, a, P_z)$ (color markers at $1/\ell = 0$) based on statistical fits in a chosen window of $\ell/a$ (bands).\label{fig:lextrap-main}}]{
 \includegraphics{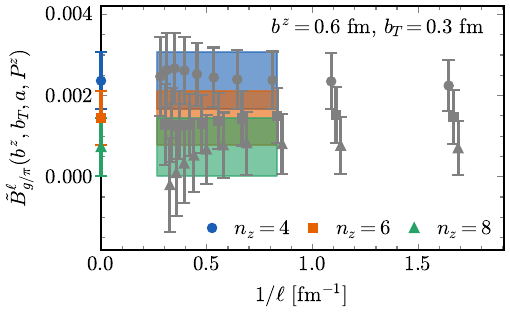}}
 \hfill \protect \\
 \subfloat[{$\tilde{B}_{g/\pi}(b^z, b_T, a, P_z)$ as functions of $b^z$, with filled (unfilled) markers representing even (odd) $b^z/a$, and opaque (transparent) colors indicating results included in (excluded from) Fourier transformations defined in Eq.~\eqref{eq:ft-compute}.\label{fig:analysis-pos-extrap}}]{
\includegraphics{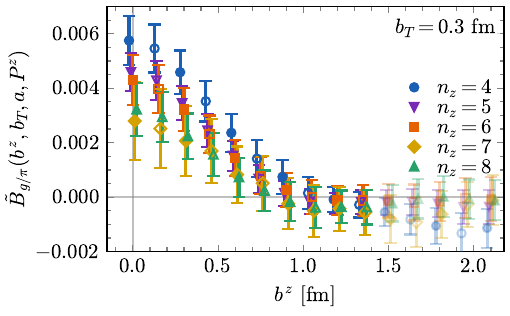}}
 \caption{Position-space functions $\tilde{B}^{\ell}_{g/\pi}(b^z, b_T, a, P_z)$ and their $\ell/a \to \infty$ extrapolations $\tilde{B}_{g/\pi}(b^z, b_T, a, P_z)$ defined in \cref{eq:matrix-element,eq:ell-extrap-ansatz}, respectively, at select $b_T$ and $P_z = 2\pi n_z/L$.
 \label{fig:analysis-pos}
}
\end{figure}
\par The $\ell$-dependent functions 
${B}^{\ell}_{g/\pi}(b^z, b_T, a, P_z)$ in Eq.~\eqref{eq:matrix-element} are obtained by combining the 
results for $\tilde{\Omega}^{\ell}_{g/\pi}(b^z, b_T, a, P_z)$, $E^\pi_{P_z}$ and
${{\tilde{\mathcal{Z}}}}^{2\ell}_g(b_T, a)$ at the outer level of correlated nested bootstrap samples, after $Z$-factor Cullum-Willoughby (ZCW) filtering and median averaging over the inner level~\cite{Wagman:2024rid,Hackett:2024xnx,Hackett:2024nbe}.
Examples of the dependence ${B}^{\ell}_{g/\pi}(b^z, b_T, a, P_z)$ on $\ell$ is illustrated in  Fig.~\ref{fig:lextrap-main}.
As expected for gluon distributions, all numerical results are real-valued and symmetric in $b^z \to -b^z$ within machine precision, and are therefore illustrated for $b^z \geq 0$ throughout this work.
\par {\it {Extrapolation $\ell/a \to \infty$}.}\Emdash{}The $\ell$-extrapolated functions in Eq.~\eqref{eq:ft} are extracted with bootstrap-level correlated fits based on the expected functional form~\cite{Ebert:2019okf}
\begin{equation}
\label{eq:ell-extrap-ansatz}
  \tilde{B}^{\ell}_{g/\pi}(b^z, b_T, a, P_z) =
  \tilde{B}_{g/\pi}(b^z, b_T, a, P_z) + \ldots\thinspace,
\end{equation}
where the ellipsis denotes $\ell$-dependent corrections.
Final results are obtained using fits to a constant excluding the corrections, separately at each $|b^z|/a \leq 14$, $b_T$, and $P_z$, in the range $\ell \in [\ell_{\text{min}},\ell_{\text{max}}]$ where $\ell_{\text{min}} = \SI{1.20}{\femto\meter}$ ($\ell_{\text{min}}/a = 8$) and $\ell_{\text{max}} = L/2 =\SI{2.40}{\femto\meter}$ ($\ell_{\text{max}}/a = 16$).  
Configurations with $|b^z|/a \in \lbrace 15, 16 \rbrace$ are excluded from the fits because fewer than $3$ values of $\ell/a$ per $b^z$ are available in this range.
Variations in the fitting range and more detailed functional forms are found to yield consistent results.
Examples of $\ell$-extrapolated functions are shown in Fig.~\ref{fig:analysis-pos}.
\par {\it {Fourier transformation}.}\Emdash{}The $x$-dependent functions $\tilde{B}_{g/\pi}\left(x,b_T,a, P_z\right)$ defined in Eq.~\eqref{eq:ft} are obtained using a discrete Fourier transformation (DFT),
\begin{equation}
\label{eq:ft-compute}
\begin{aligned}
  &x \tilde{B}^{\text{DFT}}_{g/\pi}\left(x, b_T,a, P_z, b^z_{\mathrm{cut}}\right) 
    \\&\quad= \frac{P_z}{2\pi} \sum_{b^z\leq b^z_{\mathrm{cut}}} e^{i x P_z b^z} {\tilde{B}}_{g/\pi}\left(b^z, b_T,a, P_z\right)\thinspace,
\end{aligned}
\end{equation}
where $b^z_{\mathrm{cut}}$ parametrizes the truncation point.
Final results are obtained with  $b^z_{\mathrm{cut}} = \SI{1.35}{\femto\meter}$ ($b^z_{\mathrm{cut}}/a = 9$).
Within uncertainties, the final CS kernel constraints are found to be insensitive to variations in
$b^z_{\mathrm{cut}} \in [5, 10]$, and to the addition of analytic transformations in the $\lbrack b^z_{\mathrm{cut}}, \infty)$ region based on fits to several choices of asymptotic forms derived in Ref.~\cite{Ji:2026vir} and detailed in the Supplementary Material.
\begin{figure}
\centering
\includegraphics{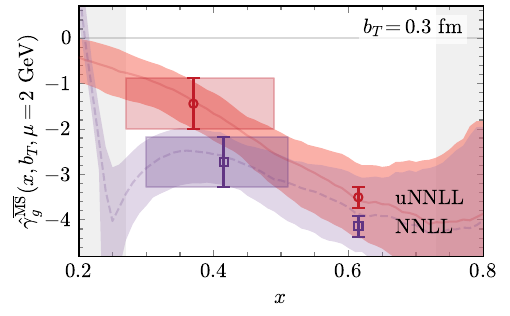}
\caption{Estimates of the gluon CS kernel $\hat{\gamma}_g(x, b_T, \mu, a)$ defined in Eq.~\eqref{eq:gammahat} as functions of $x$ for uNNLL and NNLL matching accuracies described in text.
Gray shaded areas represent the values of $x$ excluded by bounds on power corrections in Eq.~\eqref{eq:alphascut}.
Markers (horizontally offset for clarity) represent the extracted values of ${\gamma}_g(b_T, \mu)$ at fixed $b_T$.
Horizontal bands represent the $x$-ranges satisfying the goodness-of-fit criterion in Eq.~\eqref{eq:chisq-constraint} and used for each determination.
\label{fig:analysis-cs}}
\end{figure}
\par {\it {Extraction of the gluon CS kernel}.}\Emdash{}Estimates of the gluon CS kernel $\hat{\gamma}_g(x, b_T, \mu, a)$ are extracted with bootstrap-level correlated fits based on the form of $P_z$-evolution corresponding to Eq.~\eqref{eq:cs-kernel-derivative-matched} up to power corrections,
\begin{equation}
\label{eq:gammahat}
\begin{aligned}
  &\tilde{B}_{g/\pi}\left(x, b_T,a, P_z\right)   / H_g^{\text{uNNLL}}(x, b_T, \mu, P_z)
  \\ &\quad=
  c_{\gamma{\text{-fit}}}(x, b_T, a) \exp\left\lbrack\hat{\gamma}_g(x, b_T, \mu, a) \ln P_z\right\rbrack\thinspace.
\end{aligned}
\end{equation}
where $c_{\gamma{\text{-fit}}}(x, b_T, a)$ denotes a fitting parameter in addition to $\hat{\gamma}_g(x, b_T, \mu, a)$.
Analogous functional forms are used at the NNLL accuracy, with all fits performed separately at each $b_T$ and $x$, over variable $x$-ranges with uniform spacing $\Delta x = 0.01$, and over all calculated $P_z$ values.
Final results are found to be numerically stable for all $x$-ranges within $x \in [0.2, 0.8]$ and insensitive to order-of-magnitude variations in $\Delta x$.
Examples of the resulting best-fit values of $\hat{\gamma}_g(x, b_T, \mu, a)$ as a function of $x$ are illustrated in Fig.~\ref{fig:analysis-cs}. 
Within the available precision, further fits of expected power corrections, both to $\hat{\gamma}_g(x, b_T, \mu, a)$ and directly to ${\tilde{B}}^{\text{DFT}}_{g/\pi}\left(x,b_T,a, P_z\right)$, are found to be numerically unstable.
\begin{figure}
 \centering
\includegraphics{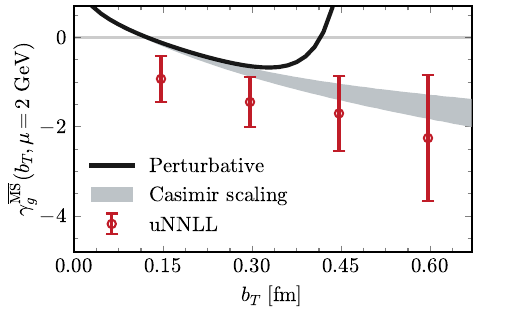}
 \caption{Final constraints on the gluon CS kernel ${\gamma}_g(b_T,\mu)$ as a function of $b_T$, extracted at uNNLL accuracy as described in text, compared with perturbative results from Ref.~\cite{Moos:2025sal} at N$^4$LL~\cite{Duhr:2022yyp,Moult:2022xzt} and the Casimir scaling of the continuum-extrapolated determination of the quark CS kernel in Ref.~\cite{Avkhadiev:2024mgd}.
 Systematic effects associated with discretization artifacts and perturbative matching are not fully controlled, and are expected to be significant for $b_T \lesssim \SI{0.30}{\femto\meter}$ ($b_T/a \lesssim 2$).\label{fig:results}}
\end{figure}
\par The final constraints on the CS kernel are extracted from $b_T$-dependent ranges $x\in [x_{\text{min}}, x_{\text{max}}] \subseteq [0.2,0.8]$ using $\hat{\gamma}_g(x, b_T, \mu, a)$ at uNNLL accuracy. 
These ranges are defined by two inequalities.
The first one is given by
\begin{equation}
\label{eq:alphascut}
\alpha^2_\text{s}(2 \bar{x} P_{z}) < \delta_{\text{p.c.}}\thinspace,
\end{equation}
where $\alpha_{\text{s}}(\mu)$ denotes the strong coupling constant at next-to-leading order (NLO) accuracy, and $\bar{x} \in \lbrace x, 1-x \rbrace$ corresponds to constraints on perturbative active-parton and nonperturbative spectator effects, respectively. 
In the final $x$-intervals, the choice $\delta_{\text{p.c.}} = 0.5$ and the smallest included $P_z \approx \SI{1.03}{\GeV}$ ($n_z = 4$) result in a $b_T$-independent constraint $x \in [0.27, 0.73]$.
The second inequality is given by
\begin{equation}
\label{eq:chisq-constraint}
    \chi^2(x, b_T) < (1+\delta_{\chi^2}) \chi^2_{\text{min}}(b_T)\thinspace,
\end{equation}
where $\chi^2(x, b_T)$ denotes the mean chi-squared statistic from fits based on Eq.~\eqref{eq:gammahat}, $\chi^2_{\text{min}}(b_T)$ denotes the corresponding minimum in $x$ within the range determined by Eq.~\eqref{eq:alphascut}, and the accepted $x$-range is restricted to a single interval containing the minimum.
In the final results, the choice $\delta_{\chi^2} = 0.5$ is applied to $\chi^2_\text{min}(b_T) \approx 1.0$ at each $b_T$, resulting in $x$-intervals asymmetric with respect to $x \to 1-x$.
The preference for asymmetric $x$-intervals by goodness-of-fit statistics is robust to variations in the extraction procedure and other analysis choices affecting the $x$-dependence of $\hat{\gamma}_g(x, b_T, \mu, a)$.
\par Within the final $x$-interval at each $b_T$, ${\gamma}_g(b_T,\mu)$ is extracted as a weighted average, with weights at each $x$ given by the ratio of the $p$-value of a bootstrap-level fit to the variance of bootstrap samples at that $x$. 
The central value is taken as the median of this average over bootstrap samples, and the associated uncertainties are defined using the empirical bootstrap confidence intervals as detailed in Ref.~\cite{Hackett:2024nbe}.
The resulting constraints at uNNLL accuracy may be compared to uNNLL and NNLL estimates $\hat{\gamma}_g(x, b_T, \mu, a)$ at $b_T = \SI{0.30}{\femto\meter}$ ($b_T/a = 2$) in Fig.~\ref{fig:analysis-cs}, with the full set of final results illustrated in Fig.~\ref{fig:results}.
\par {\it {Statistical and systematic uncertainties}.}\Emdash{}The statistical precision in the final results is comparable to that in the early-stage determination of the quark kernel in Ref.~\cite{Shanahan:2021tst}, at over an order of magnitude greater computational cost.
Both this increase and the more severe degradation of the signal-to-noise ratio in numerical results with $b_T$ are expected in the gluon case, attributable in particular to the effective doubling of Wilson line lengths in the adjoint representation.
Therefore, matching the precision and $b_T$ reach of quark-case follow-up calculations~\cite{Avkhadiev:2023poz,Avkhadiev:2024mgd} may be impractical with Wilson-line observables\Emdash{}but feasible with higher-precision TMD observables expected in the Coulomb-gauge formalism in which the relevant Wilson lines are equal to unity~\cite{Gao:2023lny,Zhao:2023ptv,Gao:2024fbh,Bollweg:2024zet,Bollweg:2025iol}, or with flow-based methods in which the variance of gluon observables may be significantly reduced~\cite{Bacchio:2023all,Catumba:2025ljd,Abbott:2026ylv,Abbott:2026vui}.
\par Several systematic effects are negligible at present precision but will grow in importance as it improves.
These include power corrections for $b_T \gtrsim \SI{0.30}{\femto\meter}$ ($b_T/a \gtrsim 2$) as illustrated in Fig.~\ref{fig:analysis-cs}, and effects arising from a finite extent of numerical results in the Wilson-line length $\ell$ and collinear separations $b^z$. 
Constraining these effects will benefit from theoretical advances in deriving  next-to-next-to-leading order (NNLO) gluon quasi-TMD matching in LaMET and physically-motivated extrapolation ans{\"a}tze such as derived in Ref.~\cite{Ji:2026vir}.
\par Other systematic effects\Emdash{}$\mathcal{O}(a)$ discretization artifacts and power corrections for $b_T \lesssim \SI{0.30}{\femto\meter}$ ($b_T/a \lesssim 2$)\Emdash{}cannot be fully quantified from the present results and are expected to be significant at the current precision level.
The expected magnitude of these effects is illustrated in Fig.~\ref{fig:analysis-cs} by the difference of the results at uNNLL and NNLL accuracies, and, {as illustrated in Fig.~\ref{fig:results}, is comparable to the difference between the final results and the perturbative determination of the kernel (up to the pole from its logarithmic resummation), as well as with the perturbatively-expected Casimir scaling of the quark kernel, $(C_\text{A}/C_{\text{F}}){\gamma}_q( b_T, \mu)$ as determined from the continuum-extrapolated $1\sigma$ uncertainty band of Ref.~\cite{Avkhadiev:2024mgd}.}
Addressing the effect of matching corrections at small $b_T$, in gluon as well as in the quark case, requires deriving the full, convolutional $b_T$-dependent matching in LaMET~\cite{Avkhadiev:2023poz}.
Control over the $\mathcal{O}(a)$ discretization artifacts requires calculation at several lattice spacings, with improvements in statistical efficiency expected for Coulomb-gauge observables~\cite{Bollweg:2024zet,Mukherjee:2024xie,Bollweg:2025iol}.
\par \textbf{Summary:}  
This work presents the first nonperturbative constraints on the gluon CS kernel ${\gamma}_g(b_T,\mu)$.
The results advance the QCD description of transverse gluon structure of hadrons, opening the door to future determinations of the gluon CS kernel with phenomenologically relevant precision directly from lattice QCD.
These determinations will test the perturbative Casimir scaling of the quark CS kernel with higher precision, may aid ongoing phenomenological studies~\cite{Echevarria:2026vca,Kishore:2022ddb,Anedda:2025bts,Kato:2024vzt,Scarpa:2019fol,Chakrabarti:2022rjr,Anedda:2026cox} and will inform future global analyses as more experimental data becomes available.
\newline
\begin{acknowledgments}
\par The QLua \cite{qlua}, QUDA \cite{Clark:2009wm,Babich:2011np,Clark:2016rdz}, and QDP-JIT \cite{6877336} software libraries were used in this work.
Data analysis used Mathematica~\cite{Mathematica}, NumPy~\cite{harris2020array}, SciPy~\cite{2020SciPy-NMeth}, and Xarray~\cite{hoyer2017xarray}, and figures were produced using Matplotlib~\cite{Hunter:2007}. 
\newline
\par We thank Christine Aidala and Renee Fatemi for helpful discussions, and the Institute for Nuclear Theory at the University of Washington for its kind hospitality and stimulating research environment.
This research was supported in part by the INT's U.S. Department of Energy grant No. DE-FG02-00ER41132.
This work was performed in part at Aspen Center for Physics, which is supported by National Science Foundation grant PHY-2210452 and a grant from the Simons Foundation (1161654, Troyer).
Argonne National Laboratory's contribution is based upon work supported by Laboratory Directed Research and Development (LDRD) funding from Argonne National Laboratory, provided by the Director, Office of Science, of the U.S. DOE under Contract No. DE-AC02-06CH11357.
This manuscript has been authored by Fermi Forward Discovery Group, LLC under Contract No. 89243024CSC000002 with the U.S. Department of Energy, Office of Science, Office of High Energy Physics.
PES is supported in part by the U.S. Department of Energy, Office of Science, Office of Nuclear Physics, under grant Contract Number DE-SC0011090 and by Early Career Award DE-SC0021006, and by Simons Foundation grant 994314 (Simons Collaboration on Confinement and QCD Strings), and has benefited from the QGT Topical Collaboration
DE-SC0023646. 
The work of YZ is supported by the U.S. Department of Energy, Office of Science, Office of Nuclear Physics through Contract No.~DE-AC02-06CH11357, and the Early Career Award through Contract No.~DE-SCL0000017.
\par This research used resources of the National Energy Research Scientific Computing Center (NERSC), a U.S. Department of Energy Office of Science User Facility operated under Contract No. DE-AC02-05CH11231, the Extreme Science and Engineering Discovery Environment (XSEDE) Bridges-2 at the Pittsburgh Supercomputing Center (PSC) through allocation TG-PHY200036, which is supported by National Science Foundation grant number ACI-1548562, facilities of the USQCD Collaboration, which are funded by the Office of Science of the U.S. Department of Energy. 
\end{acknowledgments}

\newpage
\section*{Supplementary Material}
\par This Supplementary Material (SM) collates additional results and provides further details from intermediate analysis stages to obtain the constraints on the gluon Collins-Soper (CS) kernel.
\subsection{Correlation functions}
\par The correlation functions defined in Eqs.~\eqref{eq:2pt}\myshortndash{}\eqref{eq:3pt} are computed as follows. 
On each gauge field configuration gradient-flown to $\mathfrak{t}/a^2 = 1.0$, two-point functions $C_{\text{2pt}}(t,a, \mathbf{P})$ are computed using a set of randomized source coordinates  with $16$ distinct source times $\lbrace t_0 \rbrace$ on each configuration, totaling $\num{6459392}$ measurements on the ensemble.\footnote{{In detail: on each gauge field configuration, $N^{t}_\text{src} =  16$ distinct $t_0$ coordinates are arranged on a regular grid with spacing $3a$ and a global offset sampled from $\{ 0, a, 2a \}$; on a given configuration, an equal number $ N^{x}_{\text{src}} \times N^{y}_{\text{src}} \times N^z_{\text{src}}$ of spatial source coordinates is constructed at each $t_0$ by sampling $N^z_{\text{src}}=8$ distinct $z$-planes and $N^{x}_{\text{src}}$ and $N^{y}_{\text{src}}$ distinct $x$- and  $y$-planes.
$N^{x}_{\text{src}}$ and  $N^{y}_{\text{src}}$ vary by configuration stream; for each of the three streams used, the corresponding total number of sources per configuration is given by $N_\text{src} = 3584$, $6656$, and $7168$.
}}
On the same configurations (gradient-flown to $\mathfrak{t}/a^2 = 2.0$), staple-shaped operators $O_g^\ell((\mathbf{y}, \tau), b^z, \mathbf{b}_T)$ are averaged over transverse directions $\lbrace \pm \hat{\mathbf{x}}, \pm\hat{\mathbf{y}} \rbrace$ %
and collinear orientations $\lbrace \pm \ell \rbrace$, and shifted in time for each $t_0$.
Three-point functions $C^{\ell}_{\text{3pt}}(t, \tau, b^z, b_T, a, \mom{P})$ are computed by averaging  $C_{\text{2pt}}(t,a, \mathbf{P})$ over source coordinates with equal $t_0$ on each configuration separately for forward- and backward-propagating directions in $t$; 
correlating these averages with the operator measurements configuration-by-configuration, and averaging over $\lbrace t_0 \rbrace$ and over forward and backward propagation; and subtracting the disconnected (vacuum) contributions.
\subsection{Effective energies and hadronic matrix elements\label{sec:lanczos-and-fits}}
\par Estimates of $\tilde{\Omega}^{\ell}_{g/\pi}(b^z, b_T, a, P_z)$ and $E^\pi_{P_z}$ are obtained using the Lanczos algorithm~\cite{Wagman:2024rid,Hackett:2024xnx,Ostmeyer:2024qgu,Chakraborty:2024exj,Hackett:2024nbe,Abbott:2025yhm} as follows:
\begin{enumerate}[label=(\roman*)]
    \item $C_{\text{2pt}}(t,a, \mathbf{P})$ is averaged over sources and forward- and backward propagation in $t$, excluding results for $t < t_{\text{cut}} = 4$ to avoid contact-term contamination ($t_{\text{cut}} = 6$ is confirmed to yield consistent and less statistically precise results).
    \item \label{step:lanczos-2pt} $M = \tfrac{T-t_{\mathrm{cut}}}{2}$ iterations of the algorithm are applied to $\mathrm{Re}[C^{\pi}_{\text{2pt}}(t, a, \mom{P})]$, $t \geq t_\mathrm{cut}$, with Hermitian-subspace and spurious-state filtering based on the ZCW test and nonspurious state labeling as described in Ref.~\cite{Hackett:2024nbe,Hackett:2024xnx,Abbott:2025yhm}
    For each iteration $m$, this procedure results in estimates of $E^\pi_{P_z}$ and the corresponding Ritz coefficients $P^\pi_{t}$ such that $Z_\pi(P_z) = 2E_\mom{P}^\pi \sum_{t<m} P^\pi_{t} \lbrack C_{\text{2pt}}(t, a, \mom{P}) \rbrack$~\cite{Abbott:2025yhm}.
    \item \label{step:lanczos-3pt} Estimates of $E_{P_z}^\pi$ and $P^\pi_{t}$ for each $m$ are used to obtain the corresponding estimate of the hadronic matrix element according to
    \begin{equation}
    \begin{aligned}
        &\qquad\tilde{\Omega}^{\ell}_{g/\pi}(b^z, b_T, a, P_z)
        \\ &\quad\quad=  2 E^{\pi}_{P_z} \sum_{\mathclap{\substack{\sigma, \tau \\ \sigma + \tau < m}}} P^\pi_{\sigma}  C^{\ell}_{\text{3pt}}(\sigma+\tau, \tau, b^z, b_T, a, \mom{P}) P^{\pi*}_{{\tau}}\thinspace,
    \end{aligned}
    \end{equation}
    as described further in Ref.~\cite{Hackett:2024nbe,Hackett:2024xnx,Abbott:2025yhm}.
\end{enumerate}
Steps~\ref{step:lanczos-2pt}\myshortndash{}\ref{step:lanczos-3pt} are applied to correlated nested bootstrap resampling of two- and three-point functions with $N^{\text{inner}}_{\text{b}} = N^{\text{outer}}_{\text{b}} = 200$ bootstrap samples, with central values and uncertainties at each $m$ obtained using nested median estimators and empirical bootstrap confidence intervals, respectively, as detailed further in Ref.~\cite{Hackett:2024nbe}. %
Estimates and uncertainties across all $P_z$ are observed to converge for $m \gtrsim 10$, and an average over last $N_{m} = 5$ iterations is used in the final results.
Within uncertainties, the final estimates of $\tilde{\Omega}^{\ell}_{g/\pi}(b^z, b_T, a, P_z)$ and $E^\pi_{P_z}$ are confirmed to be insensitive to variations in $N_m$ by a factor of $2$, and to variations in numerical tolerances in the state filtering procedure by an order of magnitude.
\par The Lanczos estimates of  $\tilde{\Omega}^{\ell}_{g/\pi}(b^z, b_T, a, P_z)$ and $E^\pi_{P_z}$ are also confirmed to be consistent with results from statistical fits based on spectral representations of two- and three-point correlation functions given by
\begin{equation}
\label{eq:2pt-ansatz}
\begin{aligned}
    &C_{\text{2pt}}(t, a, \mom{P})
    \\ &\qquad= \sum_{\mathfrak{n}=0}^{\infty} \frac{|Z_{\mathfrak{n}\pi}(P_z)|}{(2 E^{\mathfrak{n}\pi}_{P_z})}\left\lbrack e^{-E^{\mathfrak{n}\pi}_{P_z} t } + e^{-E^{\mathfrak{n}\pi}_{P_z} (T-t) }\right\rbrack\thinspace + \ldots
\end{aligned}
\end{equation}
and 
\begin{equation}
\label{eq:3pt-ansatz}
\begin{aligned}
    &C^{\ell}_{\text{3pt}}(t, \tau, b^z, b_T, a, \mom{P} = P_z \mathbf{\hat{z}})
    \\  &\qquad= \sum_{\mathfrak{n},\mathfrak{m}=0}^{\infty} J^\ell_{\mathfrak{n}\mathfrak{m}}(b^z, b_T, a) \frac{\sqrt{Z^*_{\mathfrak{n}}(P_z)}\sqrt{Z_{\mathfrak{m}}(P_z)}}{(2 E^{\mathfrak{n}}_{P_z})(2 E^{\mathfrak{m}}_{P_z})}
    \\ &\qquad\qquad\qquad\times  \left\lbrack e^{-E^{\mathfrak{m}}_{P_z} \tau } e^{-E^{\mathfrak{n}}_{P_z} (t-\tau) }\right\rbrack\thinspace + \ldots
\end{aligned}
\end{equation}
Here, $\mathfrak{n}$, $\mathfrak{m}$ index $\ket{\pi_{\mathfrak{n}}}$, $\ket{\pi_{\mathfrak{m}}}$: relativistically normalized eigenstates of the lattice QCD transfer matrix with the quantum numbers of the pion ordered by increasing energies $E^{\mathfrak{n}\pi}_{P_z}$, $E^{\mathfrak{m}\pi}_{P_z}$; and the ellipses denote exponentially-suppressed corrections.
The eigenstates are characterized by overlap factors $\sqrt{Z^*_{\mathfrak{n}\pi}(P_z)} = \mel{\pi_{\mathfrak{n}}(P_z)}{\chi_{\pi}^\dagger(\mathbf{0}, 0)}{0}$ and matrix elements $J^\ell_{\mathfrak{n}\mathfrak{m}}(b^z, b_T, a, P_z) = \mel{\mathfrak{n}}{O_g^\ell((\mathbf{0}, 0), b^z, \mathbf{b}_T)}{\mathfrak{m}}$.
At each $P_z$, the pion states are given by lowest-energy ($\mathfrak{n}=0$) eigenstates with $E^{0\pi}_{P_z}$, $Z_{0\pi}(P_z)$ and $J^\ell_{00}(b^z, b_T, a, P_z)$ given by $E^{\pi}_{P_z}$, $Z_{\pi}(P_z)$, and $\tilde{\Omega}^{\ell}_{g/\pi}(b^z, b_T, a, P_z)$, respectively.
\par Statistical-fit estimates of $E^\pi_{P_z}$ and $\tilde{\Omega}^{\ell}_{g/\pi}(b^z, b_T, a, P_z)$ are made as follows.
$E^\pi_{P_z}$ are extracted by fitting truncations of Eq.~\eqref{eq:2pt-ansatz}, with sets of fitting parameters $\lbrace (Z_{\mathfrak{n}\pi}(P_z),  E^{\mathfrak{n}\pi}_{P_z})  \mid 0 \leq \mathfrak{n} < N \rbrace$, to bootstrap-level numerical results using correlated $\chi^2$-minimization over a variety of fitting ranges in $t$ following a fitting procedure detailed in Refs.~\cite{Shanahan:2020zxr,NPLQCD:2020ozd}.
$\tilde{\Omega}^{\ell}_{g/\pi}(b^z, b_T, a, P_z)$ are extracted at select combinations of $\ell, b^z$ and $b_T$ using the summed ratio method~\cite{Maiani:1987by,Gusken:1989ad,Bulava:2011yz,Capitani:2012gj}.
First, a ratio of three-point to two-point correlation functions,
\begin{align}
\label{eq:ratio}
    \mathcal{R}^{\ell}(t, \tau, b^z, &b_T, a, \mom{P})
        = \frac{C^{\ell}_{\text{3pt}}(t, \tau, b^z, b_T, a, \mom{P})}{C_{\text{2pt}}(t, a, \mom{P})}\thinspace,
\end{align}
is formed.
Second, the ratio is summed over the operator insertion time $\tau$,
\begin{equation}
\label{eq:ratio-summed}
\begin{aligned}
    \mathcal{R}^{\ell}_{\Sigma}(t, b^z, b_T, a, \mom{P}) = \sum_{\mathclap{a \leq \tau \leq t-a}} \mathcal{R}^{\ell}(t, \tau, b^z, b_T, a, \mom{P})\thinspace,
\end{aligned}
\end{equation}
such that
\begin{equation}
\label{eq:ratio-summed-limit}
\begin{aligned}
    &\mathcal{R}^{\ell}_{\Sigma}(t, b^z, b_T, a, \mom{P} = P_z \mathbf{\hat{z}}) 
    \\&\quad\xrightarrow{T \gg t \gg a} t \times \frac{\tilde{\Omega}^{\ell}_{g/\pi}(b^z, b_T, a, P_z)}{2 E^\pi_{P_z}}+ \text{const.} + \ldots \thinspace
\end{aligned}
\end{equation}
Here, $``\text{const.}''$ denotes terms independent of $t$, and the ellipsis denotes excited-state contributions exponentially suppressed in $t$.
Third, the summed ratio is fit over a time window $t \in [t_{\text{start}}, t_{\text{end}}]$ using the linear form 
\begin{equation}
\begin{aligned}
&\hat{\mathcal{R}}^{\ell}_{\Sigma}(t, b^z, b_T, a, P_z) 
    \\ &\qquad= 
t \times A^\ell(b^z, b_T, a, P_z) + B^\ell(b^z, b_T, a, P_z)\thinspace,
\end{aligned}
\end{equation}
and the desired matrix element can be obtained as the combination of the slope $A^\ell(b^z, b_T, a, P_z)$ with the estimate of $E^\pi_{P_z}$ extracted as described above.
All three steps are performed at bootstrap level, separately at each $\ell$, $b^z$, and $b_T$, and jointly over all calculated $P_z$.
To avoid choosing an overly aggressive fitting window, especially at larger $P_z$ where excited-state contamination is expected to be more severe, a common-window scan is performed over all $P_z$ included in this analysis. 
For each candidate window $[t_{\text{start}}, t_{\text{end}}]$, both endpoints are varied in the range $[2a, t_{\text{max}}]$ with $t_{\text{max}} = 1.95~\text{fm}$ ($t_{\text{max}}/a = 13$), and a minimum window size $t_{\text{end}} - t_{\text{start}} > 2a$ is required. 
The same fitting window is then used for all $P_z$. 
The quality of the window is judged from the combined correlated chi-squared $\chi^2_{\text{joint}} = \sum_{P_z}\chi^2(P_z)$ with $N_{\text{dof}}^{\text{joint}}=\sum_{P_z}N_{\text{dof}}(P_z)$, where the sum runs over the momenta included in the common fit. 
Only windows
satisfying $\chi^2_{\text{joint}}/N_{\text{dof}}^{\text{joint}} < 1$ are accepted. The final result is obtained from a weighted average over the accepted fitting windows, following the procedure of Ref.~\cite{Shanahan:2020zxr,NPLQCD:2020ozd}.
\begin{figure}
  \centering
  \includegraphics{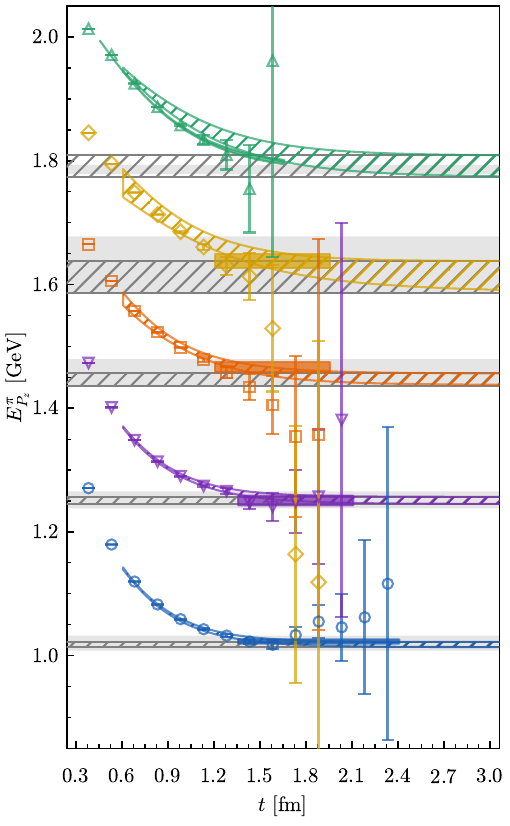}
  \caption{Effective energy functions $E^{\pi,\text{eff}}_{P_z}(t,a)$ defined in Eqs.~\eqref{eq:eff-energy}\myshortndash{}\eqref{eq:eff-energy-limit} as functions of $t$ for all calculated $P_z$ with $C_{\text{2pt}}(t, a, P_z \mom{\hat{z}})$ from numerical results (unfilled diamonds), and reconstructions of Eq.~\eqref{eq:2pt-ansatz} based on the Lanczos algorithm (hatched color bands) and highest-weight fits (solid color bands), with numerical results illustrated from $t/a=3$ to the largest $t$ in the fitting range.
  Hatched and solid gray bands represent the corresponding estimates of $E^\pi_{P_z}$ performed as described in SM Sec.~\ref{sec:lanczos-and-fits}\label{fig:e-eff}}
\end{figure}
\begin{figure}
  \centering
  \includegraphics{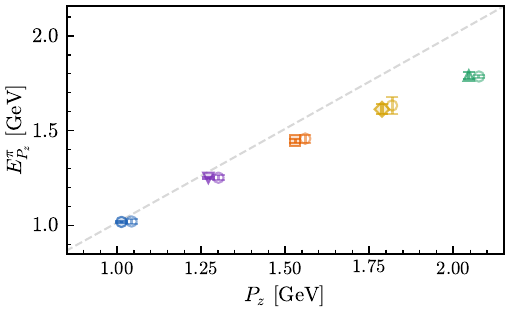}
  \caption{Continuum dispersion relation $E^{\pi,\text{cont}}_{P_z}$ defined in Eq.~\eqref{eq:continuum-dispersion} (gray line) as a function of $P_z$ compared to energy estimates $E^\pi_{P_z}$ based on Lanczos-algorithm estimates (filled markers) and statistical-fit extractions (unfilled markers, horizontally offset for clarity).\label{fig:e-disp}}
\end{figure}
\par The consistency of Lanczos and statistical-fit estimates of $E^{\pi}_{P_z}$ with each other and with numerical results for $C_{\text{2pt}}(t, a, \mom{P})$ is illustrated in Figs.~\ref{fig:e-eff}\myshortndash{}\ref{fig:e-disp}.
The comparison in Fig.~\ref{fig:e-eff} utilizes the effective energy function,
\begin{align}
\label{eq:eff-energy}
    E^{\pi,\text{eff}}_{P_z}\bigg(t+\frac{1}{2}a,\,&a\bigg) = \frac{1}{a}\log \frac{C_{\text{2pt}}(t, a, P_z \mom{\hat{z}})}{C_{\text{2pt}}(t+a, a, P_z \mom{\hat{z}})} 
    \\ &\xrightarrow{T \gg t \gg a} E^{\pi}_{P_z} + \ldots\thinspace, \label{eq:eff-energy-limit}
\end{align}
where the ellipsis denotes exponentially-suppressed corrections.
Fig.~\ref{fig:e-disp} illustrates consistency of dispersion relations between the $E^\pi_{P_z}$ estimates and $P_z$, as well as their comparison to the continuum dispersion relation defined as
\begin{equation}
\label{eq:continuum-dispersion}
    E^{\pi,\text{cont}}_{P_z} = \sqrt{\left(E^{\pi}_{P_z=0}\right)^2 + P_z^2}\thinspace,
\end{equation}
with $E^{\pi}_{P_z=0}$ based on either Lanczos or statistical-fit estimates with a statistically negligible difference, and with $a$-dependence notationally suppressed.
The relative differences between $E^{\pi,\text{cont}}_{P_z}$ and Lanczos estimates of $E^\pi_{P_z}$ with increasing $P_z = \tfrac{2\pi}{L} n_z$ for $3 \leq n_z \leq 8$ are given by $\num{0.015 \pm 0.003}$, $\num{0.028 \pm 0.004}$, $\num{0.040 \pm 0.004}$, $\num{0.072 \pm 0.006}$, $\num{0.11 \pm 0.01}$, and $\num{0.14 \pm 0.01}$\Emdash{}consistent with the linear dependence on $a$ expected for the Wilson-clover fermion action with an untuned clover-term coefficient $c_{\text{sw}}$.
Within uncertainties, the final constraints on the CS kernel are also confirmed to be insensitive to replacing $E^\pi_{P_z}$ in the definition of the normalization factor $N_{g/\pi}(a, P_z)$ in Eq.~\eqref{eq:normalization} with $E^{\pi,\text{cont}}_{P_z}$.
Further study of these discretization effects would require multiple-ensemble calculations at higher precision.
\begin{figure*}[p]
 \centering
 \subfloat[$\mathcal{R}^{\ell}(t, \tau, b^z, b_T, a, P_z\mom{\hat{z}})$ defined in Eq.~\eqref{eq:ratio} based on numerical results (color markers, horizontally offset for clarity) as functions of $\tau - t/2$ for a selection of source-sink separations $t$. 
 The color bands represent the corresponding estimates of $\tilde{\Omega}^{\ell}_{g/\pi}(b^z, b_T, a, P_z)$  based on the Lanczos algorithm (hatched) and statistical fits (solid), shown for $b^z/a = 0$ (left panels) and $b^z/a = 2$ (right panels), for boosts with $n_z=4$ (dark red, top panel) and $n_z=8$ (orange, bottom panel).\label{fig:rainbow}]{
    \includegraphics{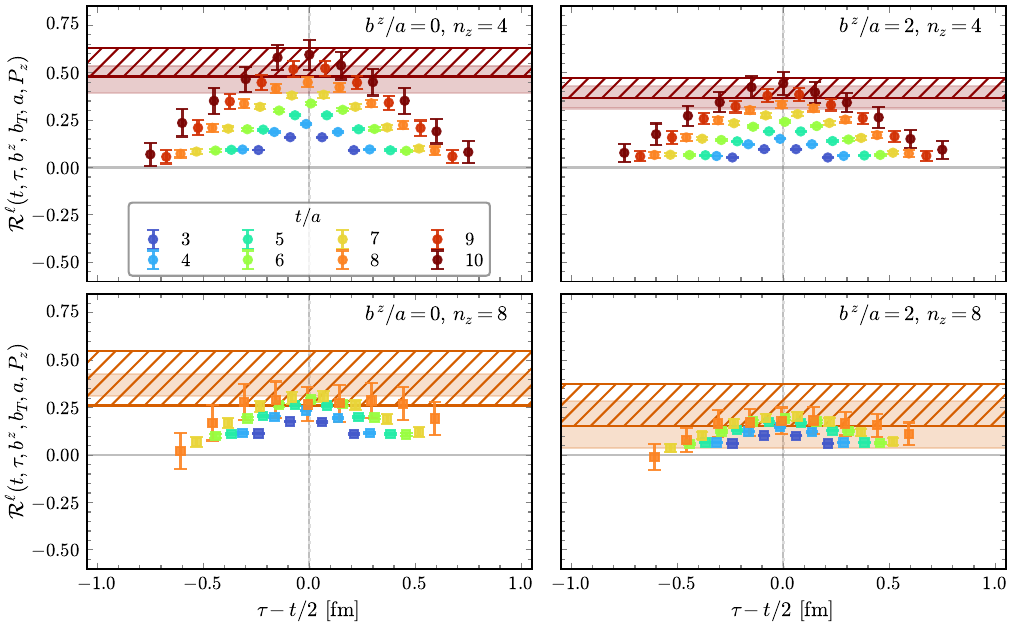}
} 
\hfill \protect \\  
\subfloat[{$\tilde{\Omega}^{\ell}_{g/\pi}(b^z, b_T, a, P_z)$ defined in Eq.~\eqref{eq:tmd-correlator} based on Lanczos-algorithm estimates (markers horizontally offset for clarity) as functions of $b^z/a$.\label{fig:omega}} %
]{
\begin{minipage}{1.0\linewidth}
        \centering
        \includegraphics{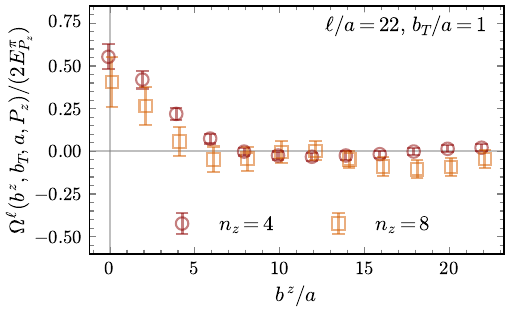}
    \end{minipage}%
} 
\caption{Lanczos and statistical-fit extractions of matrix elements $\tilde{\Omega}^{\ell}_{g/\pi}(b^z, b_T, a, P_z)$ defining the gluon quasi-TMD beam functions in Eq.~\eqref{eq:matrix-element}, performed as described in SM \cref{sec:lanczos-and-fits} at $\ell/a = 22$,  $b_T/a=1$, and two choices of boosts $P_z = 2\pi n_z/L$, compared for select $b^z/a$ to the corresponding correlation-function ratios $\mathcal{R}^{\ell}(t, \tau, b^z, b_T, a, P_z\mom{\hat{z}})$ (\ref{fig:rainbow}) and plotted as a function of $b^z/a$  (\ref{fig:omega}).\label{fig:3pt-fits}}
\end{figure*}
\par Finally, the consistency of Lanczos and statistical-fit estimates of $\tilde{\Omega}^{\ell}_{g/\pi}(b^z, b_T, a, P_z)$ with each other and with numerical results for $\mathcal{R}^{\ell}(t, \tau, b^z, b_T, a, \mom{P})$ is illustrated in Fig.~\ref{fig:3pt-fits} by the ``rainbow plots'' based on the large-time behavior of $\mathcal{R}^{\ell}_\Sigma(t, \tau, b^z, b_T, a, \mom{P})$ in Eq.~\eqref{eq:ratio-summed-limit}.
As expected, the two methods yield consistent estimates; in practice, the Lanczos algorithm is found to result in significantly more methodologically straightforward extractions with fewer and less sensitive hyperparameters.
\subsection{Divergence subtraction\label{sec:z-factor}}
\begin{figure}[t]
 \centering
\includegraphics{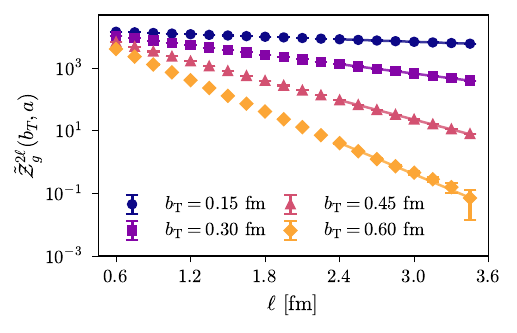}
 \caption{Numerical results for the subtraction factor $\tilde{\mathcal{Z}}^{2\ell}_g(b_T, a)$ in Eq.~\eqref{eq:matrix-element-loop} (data markers) as a function of $\ell$ for all calculated $b_T$ values, and the corresponding fits to the ansatz in Eq.~\eqref{eq:loop-exponential} (bands) performed as described in the main text and SM \cref{sec:z-factor}.\label{fig:subtraction-factor}}
\end{figure}
\par Numerical results for $\tilde{\mathcal{Z}}^{2\ell}_g(b_T, a)$ in Eq.~\eqref{eq:matrix-element-loop} are calculated beyond $\ell = L/2$ $(\ell/a = 16$) up to $\ell = \SI{3.45}{\femto\meter}$ ($\ell/a =23$).
At large $\ell$, these results are found to be consistent with the corresponding fits based on the expected functional form~\cite{Ji:2018hvs}
\begin{equation}
\begin{aligned}
\label{eq:loop-exponential}
    &{{\tilde{\mathcal{Z}}}}^{2\ell}_{g}(b_T, a) = c_{{\mathcal{Z}}}^{(1)} (b_T, a) \exp\left\lbrack- c_{{\mathcal{Z}}}^{(2)}(b_T, a)\ell \right\rbrack + \ldots\thinspace,
\end{aligned}
\end{equation}
where $c_{{\mathcal{Z}}}^{(1)}(b_T, a)$ and $c_{{\mathcal{Z}}}^{(2)} (b_T, a)$ denote free parameters, and the ellipsis denotes $\ell$-dependent corrections excluded from the fitting function.
Precisely, good agreement is observed for $\ell \gtrsim \SI{2}{\femto\meter}$ ($\ell/a \gtrsim 14$) as quantified by reduced chi-square statistic $\chi^2_\text{red} \approx 1.0$ for all $b_T/a$ and illustrated in Fig.~\ref{fig:subtraction-factor}.
Tension at smaller values of $\ell/a$ with $\chi^2_\text{red} \gtrsim 10.0$ indicates that the contribution of $\ell$-dependent power corrections not included in the functional form is significant in that range of $\ell/a$ relative to the statistical precision. 
Note that a significantly smaller contribution of $\ell$-dependent power corrections is found in the beam function $\tilde{B}^{\ell}_{g/\pi}(b^z, b_T, a, P_z)$ after dividing by $\tilde{\mathcal{Z}}^{2\ell}_g(b_T, a)$ according to \cref{eq:matrix-element}, as indicated by the results of $\ell/a \to \infty$ extrapolations described in SM \cref{sec:ell-extrap}. 
\subsection{Extrapolating \texorpdfstring{$\ell \to \infty$}{l to infinity}\label{sec:ell-extrap}}
\begin{figure}[t]
 \centering
 \includegraphics{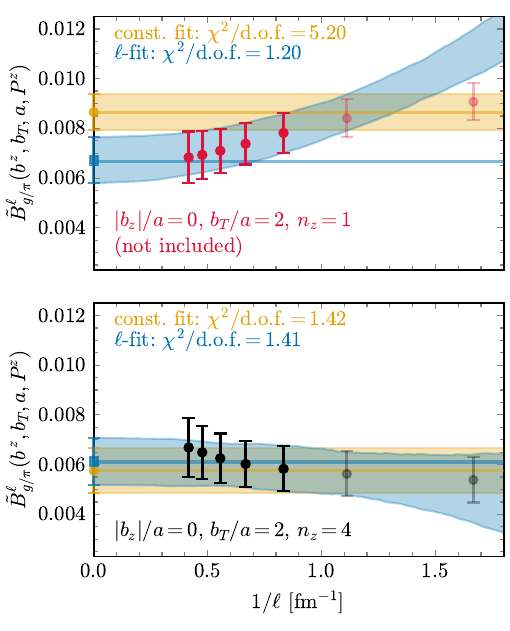}
 \caption{A comparison of fit results with two functional forms used in the $\ell/a \to \infty$ extrapolation, illustrated as functions of $1/\ell$ for select $b_T$ and $P_z$ in the $|b^z| =0$ peak of the quasi-TMD beam function with color bands, horizontal lines, and markers at $1/\ell = 0$: the constant fit based on Eq.~\eqref{eq:ell-extrap-ansatz} used in the final results as described in the main text, and the $\ell$-dependent fit in Eq.~\eqref{eq:ell-extrap-ansatz} as detailed in SM Sec.~\ref{sec:ell-extrap}.
 Numerical results included in (excluded from) the fits are illustrated with opaque (transparent) markers. 
 Results for $n_z = 1$ (red) are not included in the extraction of CS kernel constraints.
 \label{fig:lextrap}}
\end{figure}
\begin{figure*}[p]
 \centering
 \includegraphics{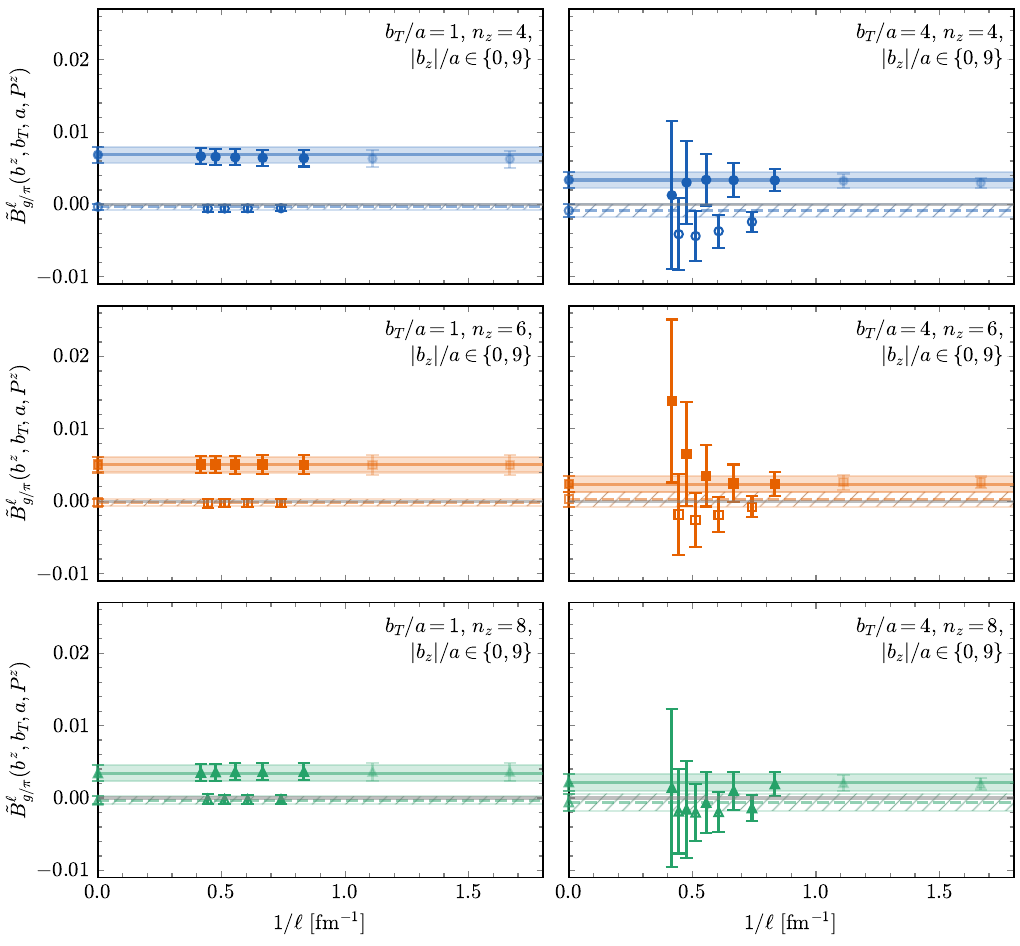}
 \caption{Constant-fit $\ell/a \to \infty$ extrapolation based on Eq.~\eqref{eq:ell-extrap-ansatz} as functions of $1/\ell$, illustrated for momenta $n_z \in \lbrace 4,6,8 \rbrace$ at the smallest $b_T=\SI{0.15}{\femto\meter}$ (left panels) and largest $b_T=\SI{0.60}{\femto\meter}$ (right panels) included in the analysis.
 Fit results at $|b^z|/a = 0$ ($|b^z|/a = 9$) are represented by solid (hatched) bands and solid (dashed) horizontal lines with filled (unfilled) markers at $1/\ell = 0$.
 The corresponding numerical results are represented by filled (unfilled) markers at finite $1/\ell$; markers for results excluded from the fit range are transparent.\label{fig:lextrap-2}}
\end{figure*}
\par A more detailed functional form used to perform the $\ell/a \to \infty$ extrapolation of ${B}^{\ell}_{g/\pi}(b^z, b_T, a, P_z)$ based on Eq.~\eqref{eq:ell-extrap-ansatz} is given by 
\begin{equation}
\label{eq:ell-fit}
\begin{aligned}
  &\tilde{B}^{\ell\text{-fit}}_{g/\pi}(b^z, b_T, a, P_z) \\ &\quad= 
  {\tilde{B}}_{g/\pi}(b^z, b_T, a, P_z)
  + \frac{c_{{\ell\text{-fit}}}(b^z, b_T, a, P_z)}{(\ell/a)^2}\thinspace,
\end{aligned}
\end{equation}
where $c_{\ell\text{-fit}}(b^z, b_T, a, P_z)$ denotes the additional fitting parameter.
As the constant-fit form included in the final analysis, Eq.~\eqref{eq:ell-fit} is fit at bootstrap level using correlated linear $\chi^2$-minimization separately at each $|b^z|$, $b_T$, and $P_z$.
\par The comparison of fits with both functional forms reveals negligible $\ell$-dependent corrections for all values of $\ell$ and $P_z$ included in the analysis, as illustrated for select matrix elements in Fig.~\ref{fig:lextrap}.
Precisely, for all $P_z$ values included in the analysis, the more detailed fit form yields consistent results with comparable values of reduced $\chi^2$ over all considered fitting ranges in $\ell$. 
However, significant improvements in the goodness-of-fit statistic are observed with matrix elements analogously extracted at lower, excluded values of $P_z \lesssim \SI{300}{\MeV}$ ($n_z = 1$) for $b_T=\SI{0.15}{\femto\meter}$ in the fitting range with $\ell_{\text{min}} \leq \SI{1.20}{\femto\meter}$ ($\ell_{\text{min}}/a \leq 8$).
In the calculations of the quark TMD matrix elements, suppression of $\ell$-dependent corrections at comparable values of $\ell$ and boosts were reported in Refs.~\cite{LPC:2022ibr,Tan:2025ofx}.
\par The results of this comparison indicate that $\ell$-dependent corrections are negligible within uncertainties for all staple lengths $\ell$ and momenta $P_z$ included in the analysis, and that the choice of $\ell_{\text{min}}$ in the final results is conservative.
Larger values of  $\ell_{\text{min}}$ in statistical fits are found to yield consistent results, with statistical uncertainties that grow severalfold larger as $\ell_{\text{min}} \to L/2$.
Consistent with this interpretation, the $\ell$-dependence of numerical results at $\ell \sim L$ illustrated in the lower panel of Fig.~\ref{fig:lextrap} and in Fig.~\ref{fig:lextrap-2} may be attributed to correlated fluctuations or finite-volume effects~\cite{Briceno:2018lfj} rather than $\ell$-dependent corrections.
\subsection{Fourier transformation \label{sec:ft}}
\begin{figure*}[t]
 \centering
 \includegraphics{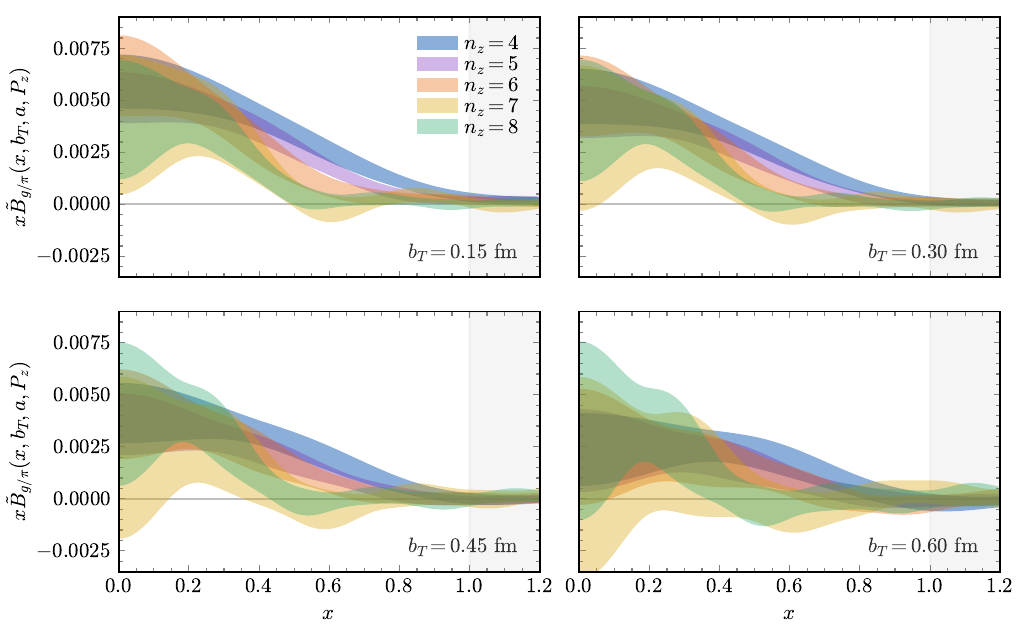}
 \caption{A complete set of LO Fourier-transformed quasi-TMD beam functions $\tilde{B}_{g/\pi}(x, b_T, a, P_z)$ defined in Eq.~\eqref{eq:ft} and implemented as DFTs $\tilde{B}^{\text{DFT}}_{g/\pi}(x, b_T,a, P_z, b^z_{\mathrm{cut}})$ with $b^z_{\text{cut}} =  9$ following Eq.~\eqref{eq:ft-compute} and the procedure described in the main text. 
 The gray shaded area represents the unphysical region $x>1$.\label{fig:x-space-compare}}
\end{figure*}
\begin{figure*}[p]
    \centering
    \subfloat[{\label{fig:ft-method}}]{\includegraphics{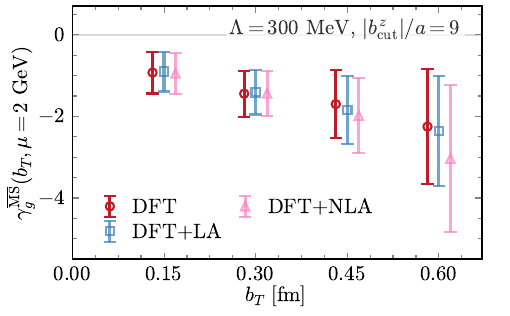}}
    \subfloat[{\label{fig:dft-model-comparison}}]{\includegraphics{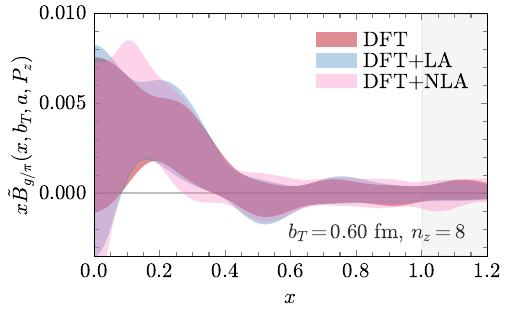}}
    \hfill \protect \\  
    \subfloat[{\label{fig:ft-lambda}}]{\includegraphics{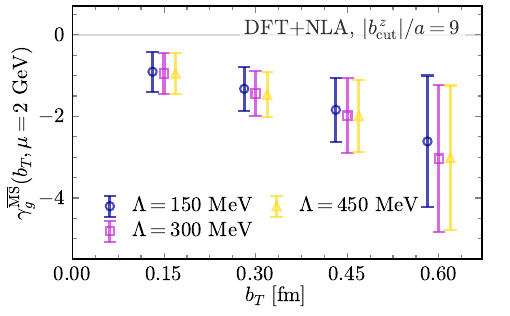}}
    \subfloat[{\label{fig:extrap-comparison}}]{\includegraphics{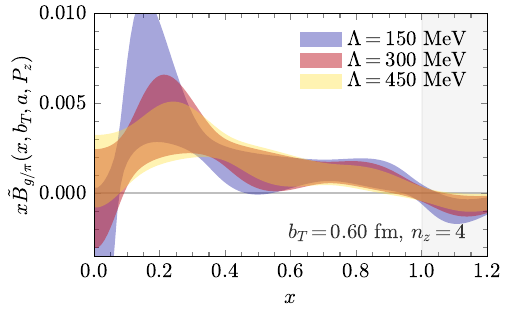}}
    \hfill \protect \\  
    \subfloat[{\parbox[t]{0.90\linewidth}{\noindent\justifying{}}
    \label{fig:final-bzcut}}]{\includegraphics{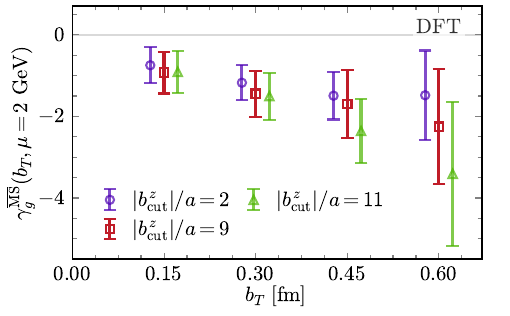}}
    \subfloat[{\parbox[t]{0.90\linewidth} {\noindent\justifying{}}
    \label{fig:x-dep-bzcut}
    }]{\includegraphics{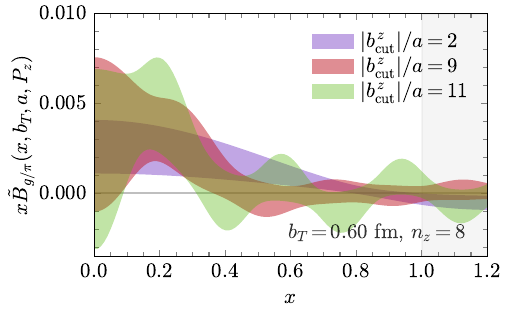}} 
    \caption{Effects of several modeling choices in implementing the Fourier transformation (FT) of \cref{eq:ft} on the final CS kernel constraints as a function of $b_T$ (left panels, markers horizontally offset for clarity) and at an intermediate analysis stage in the LO quasi-TMD beam function ${\tilde{B}}_{g/\pi}\left(x, b_T,a, P_z\right)$ as a function of $x$ (right panels) for the choice of $b_T$ and $P_z = 2 \pi n_z / L$ where the effects are most significant, with the gray shaded area representing the unphysical region $x>1$.
    The illustrated modeling choices, detailed in the main text and SM \cref{sec:ft}, comprise: LA and NLA functional forms of \cref{eq:ft-extrap-la,eq:ft-extrap-nla}, used in the analytic FT of $b^z$-tails (\cref{fig:ft-method,fig:dft-model-comparison}) and not included in the final analysis; $\Lambda$, used to set the rate of exponential decay in the functional forms and shown for the NLA form where its effects are more significant (\cref{fig:ft-method,fig:dft-model-comparison}); and $b^z_{\text{cut}}$, used to set the truncation point in the discrete FT in \cref{eq:ft-compute}, with $|b^z_{\text{cut}}|/a = 9$ used in the final results (\cref{fig:final-bzcut,fig:x-dep-bzcut}).
    }
\end{figure*}
\par This section contains \cref{fig:x-space-compare} illustrating a complete set of Fourier-transformed quasi-TMD beam functions $\tilde{B}_{g/\pi}(x, b_T, a, P_z)$ implemented as DFTs $\tilde{B}^{\text{DFT}}_{g/\pi}(x, b_T,a, P_z, b^z_{\mathrm{cut}})$ following the procedure described in the main text.
A description of the sensitivity of the analysis to alternative choices in the Fourier transformation procedure is also provided below.
\par The DFT-based transformation in Eq.~\ref{eq:ft-compute} may be extended to include the analytic transformations of the extrapolated $b^z/a \to \infty$ tails according to 
\begin{equation}
\label{eq:ft-compute-ansatz}
\begin{aligned}
  &x {\tilde{B}}_{g/\pi}\left(x,b_T,a, P_z\right) 
    \\&\quad= x \tilde{B}^{\text{DFT}}_{g/\pi}\left(x, b_T,a, P_z, b^z_{\mathrm{cut}}\right)   
    \\&\qquad + 2  \int_{b^z_{\text{cut}}}^{\infty} \frac{P^z \dd{b^z}}{2\pi}e^{i x (P_z b^z)} \tilde{B}^{\text{fit}}_{g/\pi}\left(b^z ,b_T,a, P_z\right) \thinspace,
\end{aligned}
\end{equation}
where $b^z_{\text{cut}} = \SI{1.35}{\femto\meter}$ ($b^z_{\text{cut}}/a = 9$) is used as in the final analysis, and $\tilde{B}^{\text{fit}}_{g/\pi}(b^z ,b_T,a, P_z)$ denotes the chosen asymptotic form.
This form is fit at bootstrap level using correlated nonlinear $\chi^2$-minimization to numerical results separately in the range $[b^z_{\text{ini}}, L/2]$ where $b^z_{\text{ini}} \leq b^z_{\text{cut}}$, separately at each $b_T$ and $P_z$.
\par Two choices of asymptotic forms for $\tilde{B}^{\text{fit}}_{g/\pi}(b^z ,b_T,a, P_z)$ derived in Ref.~\cite{Ji:2026vir} are considered here.
The first form is given by the leading-asymptotic (LA) ansatz
\begin{equation}
\label{eq:ft-extrap-la}
\begin{aligned}
  \tilde{B}^{\text{LA}}_{g/\pi}\left(b^z, b_T,a, P_z\right) =
  c^{(1)}_{{\text{LA}}}(b_T, a, P^z) |b^z| e^{-\Lambda |b^z|}\thinspace,
\end{aligned}
\end{equation}
with a single free parameter $c^{(1)}_{{\text{LA}}}$.
The second form is given by the next-to-leading-asymptotic (NLA) ansatz

\begin{equation}
\label{eq:ft-extrap-nla}
\begin{aligned}
  &\tilde{B}^{\text{NLA}}_{g/\pi}\left(b^z, b_T,a, P_z\right) 
  \\ &\quad=
  e^{-\Lambda |b^z|} \Big[c^{(1)}_{{\text{NLA}}}(b_T, a, P^z) |b^z| + c^{(2)}_{{\text{NLA}}}(b_T, a, P^z)
  \\  &\qquad\qquad\qquad+ 2 c^{(3)}_{{\text{NLA}}}(b_T, a, P^z)  \\  &\qquad\qquad\qquad\quad\;\,\times \cos\big(\phi_{{\text{NLA}}}(b_T, a, P^z) - b^z P_z\big) \Big] \thinspace,
\end{aligned}
\end{equation}
with four free parameters $c^{(1)}_{{\text{NLA}}}$,  $c^{(2)}_{{\text{NLA}}}$, $c^{(3)}_{{\text{NLA}}}$, and $\phi_{{\text{NLA}}}$.
In both forms, $\Lambda = \SI{300}{\MeV}$ is chosen as a fixed characteristic QCD scale governing the exponential decay, and its inclusion in the set of fitting parameters is found to result in numerically unstable fits.
In terms of goodness of fit, the best reduced $\chi^2 \approx 1.0$ is found to result for the choice of $b^z_{\text{ini}} = b^z_{\text{cut}} - 1$ in the LA fit, and for $b^z_{\text{ini}} = b^z_{\text{cut}} - 3$ in the NLA fit.
However, the NLA model is not found to be preferred by the Akaike Information Criterion (AIC)~\cite{AkaikeAIC} for any choice of the fitting range following the procedure detailed in Ref.~\cite{Shanahan:2020zxr,NPLQCD:2020ozd} with a threshold $\Delta_{\text{AIC}} < 2 N_{\text{dof}}$.
\par The addition of the analytic transformation to the DFT at fixed $b^z_{\text{cut}}/a=9$ and the associated choice of $\Lambda$ are found to have a negligible effect on the final CS kernel constraints at the current precision level.
Fig.~\ref{fig:ft-method} illustrates the effect of the modeling choice between a purely DFT-based extraction and one with either an LA-based (``DFT+LA'') or an NLA-based (``DFT+NLA'') analytic contribution.
Among the final results, the $b_T=\SI{0.60}{\femto\meter}$ ($b_T/a=4$) constraint for the DFT+NLA implementation exhibits the largest shift relative to others, but is still consistent within $1\sigma$.
As illustrated in Fig.~\ref{fig:dft-model-comparison}, this shift may be explained by the numerical ringing artifacts in the DFT+NLA transformation for the smallest $n_z = 4$ ($P_z = \SI{1.03}{\GeV})$. 
\cref{fig:ft-lambda,fig:extrap-comparison} illustrate the effect of varying $\Lambda \in [150, 300, 450]\text{ MeV}$ in the DFT+NLA transformation on the final result and the $n_z=8$, $b_T/a=4$ configuration, respectively; analogous variations in the DFT+LA model are found to be less significant.
\par The choice of $b^z_\text{cut}/a$, on the other hand, is found to have a relatively larger effect on the final CS kernel constraints, especially for $b_T/a = 4$ ($b_T = \SI{0.60}{\femto\meter}$).
This effect is illustrated in Fig.~\ref{fig:final-bzcut} for the choices of $b^z_{\text{cut}} = \SI{0.30}{\femto\meter}$ ($b^z_{\text{cut}}/a = 2$) and $b^z_{\text{cut}} = \SI{1.65}{\femto\meter}$  ($b^z_{\text{cut}}/a = 11$).
As illustrated in Fig.~\ref{fig:x-dep-bzcut}, the shift in the final results with $b^z_{\text{cut}}/a = 2$ is associated with the reduced rate of decay of the Fourier-transformed quasi-TMD beam functions in $x$, while the shift with $b^z_{\text{cut}}/a = 11$ is associated with numerical ringing artifacts in the Fourier transforms.
In turn, the ringing artifacts result from discontinuities in the $\ell$-extrapolated numerical results at large $|b^z|$, and may be attributed to correlated fluctuations and finite-volume effects in numerical results with $|b^z| \sim L$~\cite{Briceno:2018lfj}, as illustrated in \cref{fig:lextrap,fig:lextrap-2} and discussed in SM \cref{sec:ell-extrap}.
\subsection{Constraints on \texorpdfstring{$x$}{x}-intervals\label{sec:x-cuts}}
\par This section further details the inequalities used to define the fitting windows in $x$ from which the final constraints on the CS kernel are extracted using the estimates $\hat{\gamma}_g(x, b_T, \mu, a)$ in Eq.~\eqref{eq:gammahat}. 
\par
\begin{figure*}[p]
 \centering
 \subfloat[{Constraints on $x$ in Eq.~\eqref{eq:alphascut} via $\alpha_\text{s}^{\text{NLO}}(\mu)$ as defined in Eqs.~\eqref{eq:alpha-s}\myshortndash{}\eqref{eq:alpha-s-x} with $\mu = 2 x P_z$ (solid lines) and $\mu = 2 (1-x) P_z$ (dashed lines). The gray dotted line represents the bound $\delta_{\text{p.c.}}$ in the inequality. }\label{fig:as-cut}]{
 \begin{minipage}{1.0\linewidth}
        \centering
        \includegraphics{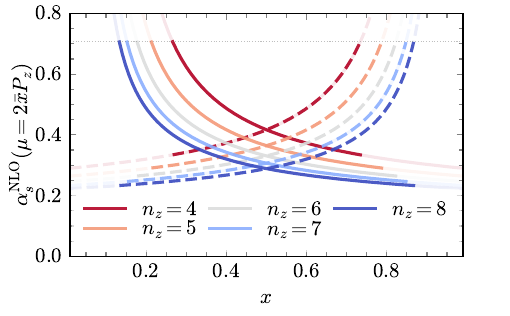}
 \end{minipage}}
 \hfill \protect \\  
 \subfloat[{Constraints on $x$ using goodness-fit statistics in Eq.~\eqref{eq:chisq-constraint} for the values of $b_T$ used in the calculation at NNLL and uNNLL matching accuracies described in the main text. 
 Gray shaded areas represent the regions of $x$ excluded by Eq.~\eqref{eq:alphascut} and illustrated in \cref{fig:as-cut} for $n_z = 4$.\label{fig:chisq-cut}}]{\includegraphics{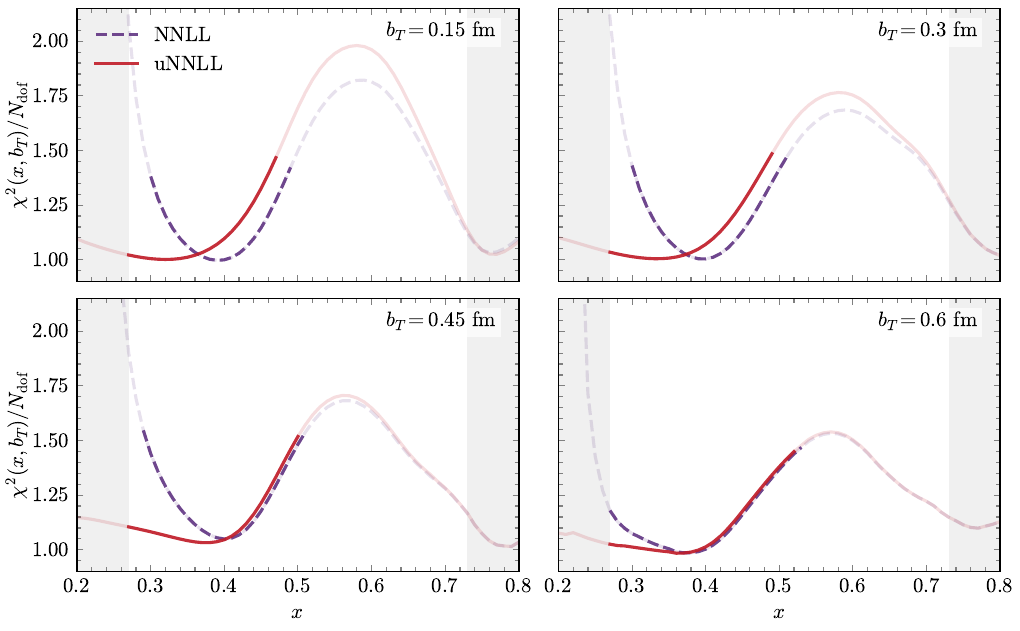}}
 \caption{Constraints on $b_T$-independent power corrections (\ref{fig:as-cut}) and goodness-of-fit statistics  (\ref{fig:chisq-cut}) defining the $x$-regions from which the final CS kernel results are extracted as described in the main text. 
 Vibrant and pale colors represent values inside and outside the bounds, respectively.\label{fig:cuts}}
\end{figure*}

 Estimates of $b_T$-independent corrections in Eq.~\eqref{eq:alphascut} utilize the strong-coupling constant calculated as
\begin{align}
\label{eq:alpha-s}
    \alpha_\text{s}^{\text{NLO}}(\mu) &= \frac{\alpha_\text{s}(\mu_0)}{X(\mu, \mu_0)}\left\lbrack1 - \frac{\alpha_\text{s}(\mu_0)}{4\pi}\frac{\beta_1}{\beta_0}\frac{\log X(\mu, \mu_0)}{X(\mu, \mu_0)}\right\rbrack\thinspace,
\\ \label{eq:alpha-s-x}
    X(\mu, \mu_0) &= 1 + \frac{\alpha_\text{s}(\mu_0)}{4\pi} \beta_0 \log\frac{\mu^2}{\mu_0^2}\thinspace,
\end{align}
where $\alpha_\text{s}(\mu_0 = \SI{2}{\GeV}) \approx \num{0.293}$ is determined as prescribed in Ref.~\cite{Bethke:2009jm}, and $\beta_n$ denotes the $n$-th order coefficient in the perturbative expansion of the QCD $\beta$-function such that
\begin{align}
    \label{eq:beta}
	&\beta_0 = \frac{11}{3}C_\text{A} - \frac{4}{3}T_\text{F} n_\text{f}\thinspace, \\
	&\beta_1 = \frac{34}{3}C_\text{A}^2 - \left(\frac{20}{3}C_\text{A} + 4C_\text{F}\right) T_\text{F} n_\text{f}\thinspace,
\end{align}
where the quadratic Casimir invariants in fundamental and adjoint representations are given by $C_{\text{F}} = 4/3$ and $C_{\text{A}}=N_{\text{c}}=3$, respectively; $T_\text{F} = 1/2$ denotes the Dynkin index in the fundamental representation; and $n_\text{f} = 4$ is chosen as the appropriate number of active quark flavors for the calculation.
The corresponding values of $\alpha_\text{s}^{\text{NLO}}(2\bar{x} P_z)$ and the regions of $x$ where the inequality is satisfied are illustrated in Fig.~\ref{fig:cuts}.
\par The additional constraints on the $x$-intervals based on goodness-of-fit statistics in Eq.~\eqref{eq:chisq-constraint} are illustrated in Fig.~\ref{fig:cuts}.
The final results are found to be insensitive to variations in $\delta_{\chi^2} \lesssim 0.6$. 
Constraints based on Eq.~\eqref{eq:alphascut} alone lead to $x$-intervals symmetric with respect to $x \to 1-x$.
Such symmetric intervals result in a negative shift of final CS kernel constraints within $1\sigma$ for $b_T \lesssim \SI{0.3}{\femto\meter}$ ($b_T/a \lesssim 2$), and are insensitive at the current precision level to the exact choice in $x \in [0.2, 0.8]$ controlled by variations in $\delta_{\text{p.c.}}$.
The difference in extractions based on symmetric and asymmetric $x$-intervals may reflect a similar asymmetry in the $x$-region where power corrections, including $b_T$-dependent effects not captured at uNNLL accuracy, are minimized for quasi-TMD beam functions.
\subsection{LaMET matching}
\par The LaMET matching kernels used in this analysis are defined as
\begin{equation}
\label{eq:matching-kernel-coeff}
    H_{g,nn^\prime}(\mu, x, P_z) 
        = C_{g,n}(\mu, xP_z) \times C_{g,n^\prime}(\mu, xP_z)\thinspace,
\end{equation}
where $H_{g,nn^\prime}(\mu, x, P_z)$ denotes the kernel, $C_{g,n}(\mu, xP_z)$ denotes the corresponding LaMET matching coefficient for a gluon quasi-TMD beam function such that
at LO $C_{g,n}^{\text{LO}}(\mu, xP_z)=1$, and $n$, $n^\prime$ correspond to the Lorentz indices in the gluon quasi-TMD operator such that $n=n^\prime = t$ for the operator defined in Eq.~\eqref{eq:operator-0i0i}. 
\par At fixed order, the matching coefficients are known up to NLO~\cite{Schindler:2022eva,Zhu:2022bja}, and are given at this order by
\begin{equation}
\label{eq:nlo}
\begin{aligned}
    &C^{\text{NLO}}_{g, t}(\mu, xP_z)
        \\ &\quad= C^{\text{NLO}}_{g,+}(\mu, xP_z) + C_\text{A} \frac{\alpha_{\text{s}}(\mu)}{4\pi}\left(\log\frac{(2 xP_z)^2}{\mu^2}-2\right)\thinspace,
\end{aligned}
\end{equation}
for the operator used in this work, with $C^{\text{NLO}}_{g,+}(\mu, xP_z)$ defined in Ref.~\cite{Zhu:2022bja}.
The corresponding coefficient for the $n=z$ is given by 
\begin{equation}
\label{eq:nlo-z}
\begin{aligned}
    &C^{\text{NLO}}_{g, z}(\mu, xP_z)
        \\ &\quad= C^{\text{NLO}}_{g,+}(\mu, xP_z) - C_\text{A} \frac{\alpha_{\text{s}}(\mu)}{4\pi}\left(\log\frac{(2 xP_z)^2}{\mu^2}-2\right)\thinspace.
\end{aligned}
\end{equation}
\par The logarithmic resummation of fixed-order matching coefficients from $\mu_0$ to $\mu$ is defined as
\begin{equation}
\begin{aligned}
    &C^{\text{N}^k\text{LL}}_{g, n}(\mu, xP_z)
        \\ &\quad= C^{\text{N}^{k-1} \text{LO}}_{g,n}(\mu_0, xP_z) \exp[- K_g^{\text{N}^k\text{LO}}(\mu_0, \mu)]\thinspace,
\end{aligned}
\end{equation}
where $C^{\text{N}^k\text{LL}}_{g, n}(\mu, xP_z)$ denotes the resummed coefficient at the $k$-th order of accuracy, $K_g^{\text{N}^k\text{LO}}(\mu_0, \mu)$ denotes the resummation kernel defined in Refs.~\cite{Gaunt:2014cfa,Luo:2019bmw} through order $k=2$ (\text{NNLO}), and $C^{\text{N}^{k-1} \text{LO}}_{g,n}(\mu_0, xP_z)$ denotes the corresponding fixed-order coefficient, with LO coefficients used for resummation accuracies of both $k=1$  ($\text{NLL}$) and $k=0$ ($\text{LL}$).
\par The $b_T$-unexpanded form of fixed-order matching coefficients is defined as
\begin{equation}
\label{eq:bt-unexp}
\begin{aligned}
    &C^{\text{uN}^{k}\text{LO}}_{g, n}(\mu, xP_z)
        \\ &\quad= C^{\text{N}^k\text{LO}}_{g,n}(\mu, xP_z) + \frac{C_\text{A}}{C_\text{F}} \delta C^{\text{N}^{k}\text{LO}}(b_T, \mu, xP_z)\thinspace,
\end{aligned}
\end{equation}
where $C^{\text{N}^{k}\text{LO}}(b_T, \mu, xP_z)$ denotes the $b_T$-dependent correction preserving the multiplicative matching form in Eq.~\eqref{eq:cs-kernel-derivative-matched} and introduced originally in Ref.~\cite{Avkhadiev:2023poz} for the quark matching, and the factor of $C_\text{A}/{C_\text{F}}$ converts the correction to the present case of gluon matching.
\begin{figure}
  \centering
  \includegraphics{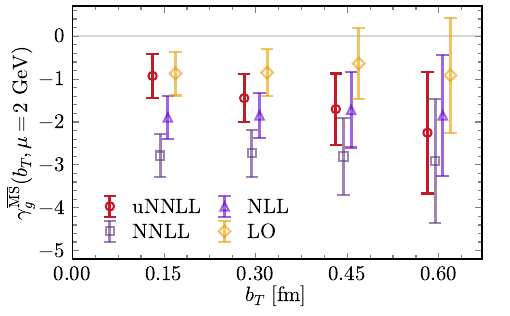}
  \caption{A comparison of CS kernel constraints as a function of $b_T$ using LaMET matching kernels at several perturbative accuracies as defined in Eqs.~\eqref{eq:unnll-2}\myshortndash{}\eqref{eq:nll} (markers horizontally offset for clarity), with final results obtained at the uNNLL accuracy.\label{fig:gammahat-final-matching}}
\end{figure}
\par Based on Eqs.~\eqref{eq:matching-kernel-coeff}\myshortndash{}\eqref{eq:bt-unexp}, the uNNLL and NNLL matching kernels used in this work are given by
\begin{equation}
\label{eq:unnll-2}
\begin{aligned}
    &H^{\text{uNNLL}}_g(x, b_T, \mu, P_z) 
        \\ &\quad= \left(C^{\text{uN}\text{LO}}_{g, t}(\mu_0, xP_z) \right)^2 \exp[- 2 K_g^{\text{NN}\text{LO}}(\mu_0, \mu)]\thinspace
\end{aligned}
\end{equation}
and
\begin{equation}
\label{eq:nnll}
\begin{aligned}
    &H^{\text{NNLL}}_g(x, b_T, \mu, P_z) 
        \\ &\quad= \left(C^{\text{N}\text{LO}}_{g, t}(\mu_0, xP_z)\right)^2\exp[- 2 K_g^{\text{NN}\text{LO}}(\mu_0, \mu)]\thinspace, 
\end{aligned}
\end{equation}
respectively, with logarithms resummed in the $\overline{\text{MS}}$ scheme from an initial scale of $\mu_0 = 2 x P_z$ to $\mu = \SI{2}{\GeV}$ following Refs.~\cite{Avkhadiev:2023poz,Avkhadiev:2024mgd}. 
For completeness, results are also obtained with matching kernels at the next-to-leading-logarithmic (NLL) accuracy, 
\begin{equation}
\label{eq:nll}
\begin{aligned}
    &H^{\text{NLL}}_g(x, b_T, \mu, P_z) 
        \\ &\quad= \left(C^{\text{LO}}_{g, t}(\mu_0, xP_z)\right)^2\exp[- 2 K_g^{\text{N}\text{LO}}(\mu_0, \mu)]\thinspace,
\end{aligned}
\end{equation}
with an identical choice of $\mu_0$.
A comparison of CS kernel constraints obtained at these accuracies as well the LO accuracy is illustrated in Fig.~\ref{fig:gammahat-final-matching}.
Compared to systematic uncertainties already indicated by the difference of final results at NNLL and uNNLL accuracies, smaller effects are found to be associated with the alternative choice of $\mu_0 = \mu/2$, as well as with variations of the initial scale in the canonical range of $\lbrack \mu_0/2, 2 \mu_0 \rbrack$ in both cases.
A more careful analysis of these effects is warranted with increased precision in numerical results, or with further progress in LaMET matching for the gluon case: the derivation of fixed-order matching coefficients at NNLO and of the full, convolutional $b_T$-dependent matching.
\par The importance of logarithmic resummation at NLO in the present analysis, and the effect of NNLL and uNNLL matching kernels on the CS kernel constraints in Eq.~\eqref{eq:cs-kernel-derivative-matched} may be expressed using the corresponding logarithmic derivatives:
\begin{equation}
\label{eq:dgamma-nlo}
\begin{aligned}
    &\delta\gamma^{\text{NLO}}_g(x, \mu, P_z) 
       \\ &\qquad =  -P_z \dv{P_z} \log H^{\text{NLO}}_g(x, \mu, P_z)\thinspace,
\end{aligned}
\end{equation}
\begin{equation}
\label{eq:dgamma-nnll}
\begin{aligned}
    &\delta\gamma^{\text{NNLL}}_g(x, \mu, P_z) 
       \\ &\qquad =  -P_z \dv{P_z} \log H^{\text{NNLL}}_g(x, \mu, P_z)\thinspace,
\end{aligned}
\end{equation}
and
\begin{equation}
\label{eq:dgamma-unnll}
\begin{aligned}
    &\delta\gamma^{\text{uNNLL}}_g(x, b_T, \mu, P_z) 
        \\ &\qquad= -P_z \dv{P_z} \log H^{\text{uNNLL}}_g(x, b_T, \mu, P_z)\thinspace,
\end{aligned}
\end{equation}
respectively.
\begin{figure*}[p]
 \centering
    \subfloat[{NLO corrections.\label{fig:dgamma-nlo}}]{
    \includegraphics{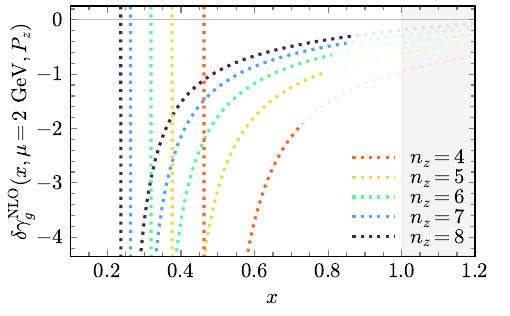}}
  \subfloat[{NNLL corrections.\label{fig:dgamma-nnll}}]{
  \includegraphics{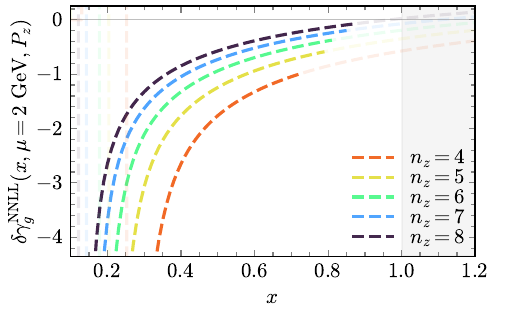}}
 \hfill \protect \\
 \subfloat[{Comparison of the $b_T$-dependent uNNLL and $b_T$-independent NNLL corrections for the values of $b_T$ used in the calculation.\label{fig:dgamma-unnll}}]{
 \includegraphics{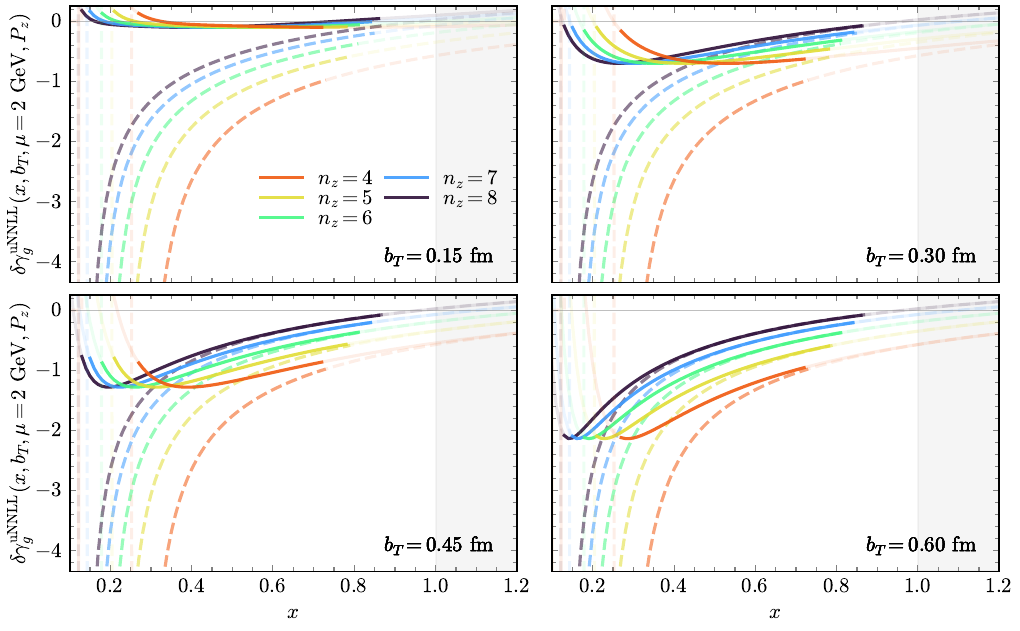}}
 \caption{Additive LaMET matching corrections to the CS kernel according to Eq.~\eqref{eq:cs-kernel-derivative-matched} as functions of $x$ for each of the calculated $P_z = 2\pi n_z / L$: $\delta\gamma^{\text{NLO}}_g(x, \mu, P_z)$ defined in Eq.~\eqref{eq:dgamma-nlo} (dotted lines in Fig.~\ref{fig:dgamma-nlo}), $\delta\gamma^{\text{NLO}}_g(x, \mu, P_z)$ defined in Eq.~\eqref{eq:dgamma-nnll} (dashed lines in Figs.~\ref{fig:dgamma-nnll}\myshortndash{}\ref{fig:dgamma-unnll}), and $\delta\gamma^{\text{uNNLL}}_g(x, b_T, \mu, P_z)$ defined in Eq.~\eqref{eq:dgamma-unnll} and used in the final results (solid lines in Fig.~\ref{fig:dgamma-unnll}).
 Vibrant and pale colors represent regions of $x$ inside and outside the constraints defined by Eq.~\eqref{eq:alphascut} and illustrated in Fig.~\ref{fig:cuts}.
 The gray shaded area represents the unphysical region $x>1$.\label{fig:dgamma}}
\end{figure*}
As illustrated in Fig.~\ref{fig:dgamma}, the logarithmic derivatives generally result in a negative, $x P_z$-dependent shift of the CS kernel constraints.
\par As illustrated in Fig.~\ref{fig:dgamma-nlo}, NLO matching corrections in Eq.~\eqref{eq:dgamma-nlo} feature singularities which, for the calculated values of $P_z$, appear in the intermediate range of $x$ illustrated in Fig.~\ref{fig:dgamma-nlo}.
These singularities preclude CS kernel extraction with NLO matching and require logarithmic resummation in this analysis.
\par As illustrated in Fig.~\ref{fig:dgamma-nnll}, NNLL matching corrections in Eq.~\eqref{eq:dgamma-nnll} result in a negative $b_T$-independent shift of the CS kernel constraints by a factor of approximately $-1$ to $-3$ in the intermediate region of $x$ for the calculated range of $P_z$. 
As expected, this shift vanishes as $x \to 1$ for $P_z \gg \Lambda$\Emdash{}since the matching coefficients for TMD beam functions do not account for the nonperturbative spectator effects\Emdash{}and exhibits rapid large-amplitude oscillations characteristic of the logarithmic enhancement in $x P_z$ as $x \to 0$; at each $P_z$, the latter region in $x$ is excluded from the analysis by the $b_T$-independent constraint in Eq.~\eqref{eq:alphascut}.
\par As illustrated in Fig.~\ref{fig:dgamma-unnll}, uNNLL matching corrections in Eq.~\eqref{eq:dgamma-unnll} result in a negative $b_T$-dependent shift which is reduced relative to that at NNLL, effectively interpolating between NNLL (the $b_T$-independent negative shift) for $b_T \gg (x P_z)^{-1}$ and LO (zero shift) for $b_T \ll x P_z$.
This small-$b_T$ limiting behavior is expected based on the multiplicative form of uNNLL matching which does not incorporate all $b_T$-dependent corrections.
The effect of additional $b_T$-dependent corrections from the full convolutional matching may be important to explain the persistent tension of the uNNLL constraints with perturbative results at small $b_T$ illustrated in Fig.~\ref{fig:results}, and requires further study beyond that described in SM \cref{sec:x-cuts}.
\subsection{Additional examples of intermediate results}
\begin{figure*}[p]
  \centering
  \subfloat[Position-space functions $\tilde{B}_{g/\pi}(b^z, b_T, a, P_z)$ as functions of $b^z$ across all calculated momenta $4 \leq n_z \leq 8$.\label{fig:position-supp}]{%
    \includegraphics{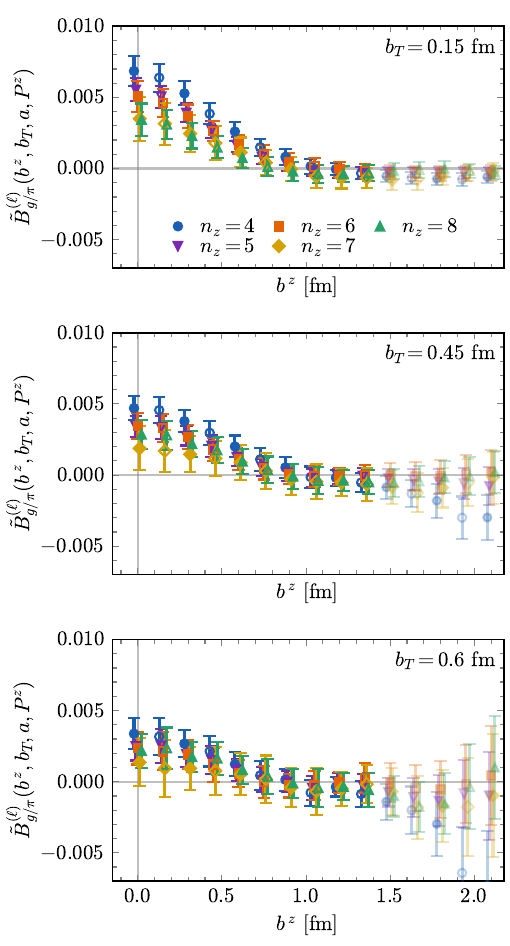}%
  }
  \hfill
  \subfloat[Estimates of the gluon CS kernel $\hat{\gamma}_g(x, b_T, \mu, a)$ defined in Eq.~\eqref{eq:gammahat} as functions of $x$.\label{fig:gammahat-supp}]{%
    \includegraphics{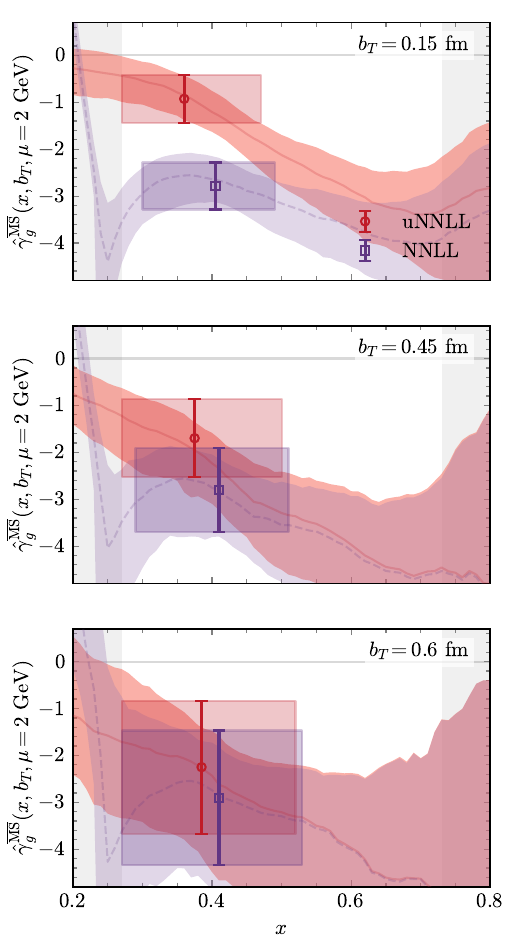}%
  }
  \caption{Intermediate-stage results for transverse separations $b_T/a \in \{1, 3, 4\}$, supplementing the corresponding $b_T/a=2$ results illustrated in Fig.~\ref{fig:analysis-pos} and Fig.~\ref{fig:analysis-cs} in the main text.}
  \label{fig:supplemental-combined}
\end{figure*}
\par This section contains figures illustrating additional examples of numerical results at intermediate analysis steps. 
Supplementing the $b_T/a = 2$ results presented in the main text, additional examples of position-space functions $\tilde{B}^{\ell}_{g/\pi}(b^z, b_T, a, P_z)$ across all studied momenta $n_z \in \{4, 5, 6, 7, 8\}$ are illustrated in Fig.~\ref{fig:position-supp} for the remaining transverse separations $b_T/a \in \lbrace 1,3,4\rbrace$. 
Supplementing Fig.~\ref{fig:analysis-cs}, the corresponding estimates of the gluon CS kernel $\hat{\gamma}_g(x, b_T, \mu, a)$ evaluated at these separations are provided in Fig.~\ref{fig:gammahat-supp}.
\subsection{Results with alternative operator definitions\label{sec:alternative-ops}}
\par Following analogous analyses, several other gluon TMD operators have been confirmed to yield constraints on the gluon CS kernel consistent with the final results presented in the main text.
Because the differences between the constraints are not significant at the current precision level, they are not used to define explicit systematic uncertainty estimates, and are not included in the main text.
For completeness, the resulting constraints and the corresponding operator definitions are detailed below.
\par In the case of unpolarized gluon TMDs, the general form of gluon TMD operators in Eq.~\eqref{eq:operator-0i0i-detail} is specialized to
\begin{equation}
\label{eq:operator-unpolarized}
\begin{aligned}
    &O_g^{ \lbrace \mu\nu \rbrace\perp \hat{e},\ell}(y, b^z, \pos{b}_T) 
    = - \tfrac{1}{2} g_{\alpha\beta}^{\perp \hat{e}} O_g^{\lbrace \mu\nu \rbrace \alpha\beta,\ell} (y, b^z, \pos{b}_T) \thinspace,
\end{aligned}
\end{equation}
where curly braces denote index symmetrization, and $g_{\alpha\beta}^{\perp\hat{e}} = g_{\alpha\beta} - \hat{e}_{\lbrace \alpha}\hat{e}_{\beta\rbrace}$ denotes the transverse metric with respect to a unit four-vector $\hat{e}$~\cite{Boussarie:2023izj}.
Two choices of $\hat{e}$ used here are given by $\hat{z}$ and  $\hat{n} = (\hat{z} + \hat{t})/\sqrt{2}$, corresponding to transverse indices $\alpha,\beta \in \lbrace x,y \rbrace$ and $\alpha,\beta \in \lbrace x,y,t \rbrace$, respectively.
To define dimensionless unpolarized gluon TMDs, matrix elements of these operators in external hadron states  such as the one in \cref{eq:matrix-element} are combined with normalization factors
\begin{equation}
\label{eq:normalization-general}
    N^{\lbrace \mu\nu \rbrace}_{g/h}(a, P)  = 1/(P^\mu P^\nu)\thinspace,
\end{equation} 
where $P^\mu$, $P^\nu$ denote the longitudinal components of the hadron.
\par The operator $O_{g}^{\ell}(y, b^z, \pos{b}_T)$ in Eq.~\eqref{eq:operator-0i0i} used to obtain the final constraints in this work is given in the notation of Eq.~\eqref{eq:operator-unpolarized} by
\begin{align}
\label{eq:0i0i-xy-xyt}
    &O_{g}^{\langle tt \rangle \perp\hat{n},\ell}(y, b^z, \pos{b}_T) 
    \nonumber \\ &\qquad=  \left(g^{t}_{\rho}g^{t}_{\sigma} - \frac{1}{3} \Delta^{tt} \Delta_{\rho\sigma} \right)O_{g}^{ \lbrace \rho\sigma \rbrace\perp \hat{n},\ell}(y, b^z, \pos{b}_T)\thinspace,
\end{align}
Here, $g_{\rho\sigma}$ denotes the components of a metric tensor, the projection operator $\Delta_{\rho\sigma} = g_{\rho\sigma} - \hat{z}_{\lbrace \rho} \hat{z}_{\sigma \rbrace}$ isolates the sum over the longitudinal components in the trace to $\rho,\sigma \in \lbrace x,y,t\rbrace$, and the label $\langle \mu\nu \rangle$ indicates a symmetric trace-subtracted combination.
For the external hadron state boosted along $P_z \mathbf{\hat{z}}$, 
the  corresponding normalization factor $N^{\lbrace\mu\nu\rbrace}_{g/h}(a, P)$ in Eq.~\eqref{eq:normalization-general} yields $N_{g/h}(a, P)$ in Eq.~\eqref{eq:normalization}.
\par Following analogous analysis steps, three additional operators  are found in this work to yield final constraints with comparable precision.
The first two correspond to the same choice of longitudinal components $\mu,\nu=t$:
\begin{align}
\label{eq:0i0i-full}
    &O_{g}^{\lbrace tt \rbrace \perp\hat{n},\ell}(y, b^z, \pos{b}_T) =  O_{g}^{\ell}(y, b^z, \pos{b}_T)\thinspace,\\
\label{eq:0i0i-xyt-xyt}
    &O_{g}^{\langle tt \rangle \perp\hat{z},\ell}(y, b^z, \pos{b}_T) 
    \nonumber \\ &\qquad= \left(g^{t}_{\rho}g^{t}_{\sigma} - \frac{1}{3} \Delta^{tt} \Delta_{\rho\sigma} \right)O_{g}^{ \lbrace \rho\sigma \rbrace\perp \hat{z},\ell}(y, b^z, \pos{b}_T)\thinspace,
\end{align}
with normalization factors $N^{\lbrace\mu\nu\rbrace}_{g/h}(a, P)$ in Eq.~\eqref{eq:normalization-general} modified accordingly.
Compared to the operator choice in the main text, the operator in Eq.\eqref {eq:0i0i-full} is found to result in smaller statistical uncertainties but higher sensitivity to finite-volume effects at large $\ell$ discussed in SM \cref{sec:ell-extrap}; the operator in Eq.\eqref {eq:0i0i-xyt-xyt}, in larger statistical uncertainties and lower sensitivity to finite-volume effects. 
A final alternative operator definition considered here corresponds to $\mu,\nu=z$:
\begin{align}
\label{eq:3i3i-xy-xyzt}
    &O_{g}^{\langle zz \rangle \perp\hat{n},\ell}(y, b^z, \pos{b}_T) 
    \nonumber \\ &\qquad=  \left(g^{z}_{\rho}g^{z}_{\sigma} - \frac{1}{4} g^{zz} g_{\rho\sigma} \right) O_{g}^{ \lbrace \rho\sigma \rbrace\perp \hat{n},\ell}(y, b^z, \pos{b}_T)\thinspace.
\end{align}
In contrast to operators in \cref{eq:0i0i-full,eq:0i0i-xy-xyt,eq:0i0i-xyt-xyt}, this operator is known to break multiplicative renormalizability by combining trace-subtracting terms where a distinct number of Lorentz components coincides with the collinear direction of the staple-shaped Wilson line ($z$) (such combinations have been shown to contain distinct ultraviolet divergences in Ref.~\cite{Zhang:2018diq}).
Omitting the trace subtraction in this operator is found to yield a prohibitively low signal-to-noise ratio in the numerical results.
\begin{figure*}[t]
  \centering
  \subfloat[LO.\label{fig:comparison-ops-lo}]{%
    \includegraphics{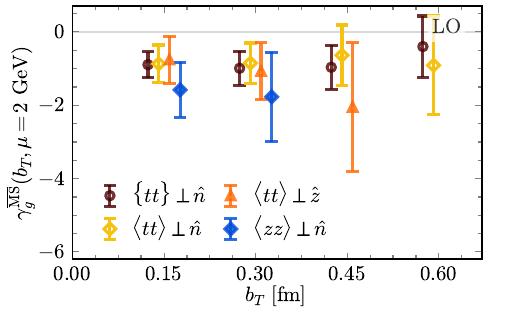}%
  }
  \hfill
  \subfloat[uNNLL.\label{fig:comparison-ops-unnll}]{%
    \includegraphics{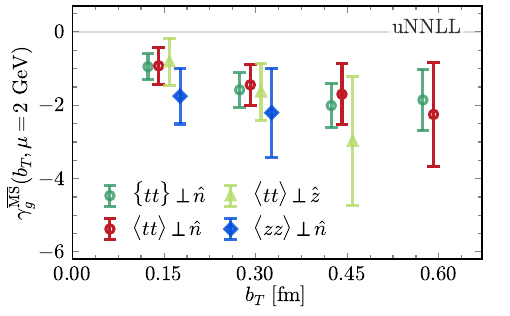}%
  }
  \caption{A comparison of CS kernel constraints as function of $b_T$ at LO (left) and uNNLL (right) LaMET matching accuracies across several definitions of gluon TMD operators in \cref{eq:0i0i-full,eq:0i0i-xy-xyt,eq:0i0i-xyt-xyt,eq:3i3i-xy-xyzt} (markers offset horizontally for clarity).
  Labels ``$\langle tt \rangle \perp \hat{n}$'' correspond to the operator in \cref{eq:0i0i-xy-xyt} used in the final results, and to identical markers in Fig.~\ref{fig:gammahat-final-matching}.
  Labels ``$\lbrace zz \rbrace \perp \hat{n}$'' correspond to the operator in \cref{eq:3i3i-xy-xyzt} which is not multiplicatively renormalizable as detailed in SM \cref{sec:alternative-ops}. 
  Results with vanishing signal-to-noise ratios are omitted.
  }
  \label{fig:comparison-ops}
\end{figure*}
\par A comparison of constraints on the gluon CS kernel from operators in \cref{eq:0i0i-full,eq:0i0i-xy-xyt,eq:0i0i-xyt-xyt,eq:3i3i-xy-xyzt} is illustrated in Fig.~\ref{fig:comparison-ops}, omitting results with vanishing signal-to-noise ratios at $b_T/a \geq 3$ for the operator in \cref{eq:0i0i-xyt-xyt} and $b_T/a \geq 2$ for the operator in \cref{eq:3i3i-xy-xyzt}.
At uNNLL matching accuracy illustrated in Fig.~\ref{fig:comparison-ops-unnll}, matching kernels for operators in \cref{eq:0i0i-full,eq:0i0i-xyt-xyt} are based on those used to obtain the main results, with a possible constant rescaling that accounts for trace subtraction and does not modify $P_z$-dependence; matching kernels for the operator in \cref{eq:3i3i-xy-xyzt} are based on the matching coefficient in \cref{eq:nlo-z}.
The resolved constraints across the operator definitions are consistent, and differences between them do not reveal systematic effects significant at the current precision level.
This suggests that the dominant systematic effects that remain not fully controlled in the present results, especially for $b_T \lesssim  \SI{0.30}{\femto\meter}$, may be common across all studied operator choices.
\end{document}